\documentclass[useAMS,usenatbib]{mnras}

\usepackage{graphicx, bm, amssymb, xcolor}
\usepackage{verbatim}
\usepackage[all]{hypcap} 
\usepackage{xspace}
\usepackage{multirow,bigdelim}
\usepackage[fleqn]{amsmath}
\usepackage{mathtools}

\usepackage{epsfig}
\usepackage{aas_macros}
\usepackage{natbib}
\usepackage{times}

\newcommand{\msun}{\mbox{M$_\odot$}}

\newcommand{\myr}{\mbox{${\rm Myr}$}}

\newcommand{\pc}{\mbox{${\rm pc}$}}
\newcommand{\kpc}{\mbox{${\rm kpc}$}}

\newcommand{\kms}{\mbox{${\rm km}~{\rm s}^{-1}$}}
\newcommand{\cmc}{\mbox{${\rm cm}^{-3}$}}

\newcommand{\dex}{\mbox{${\rm dex}$}}

\newcommand{\menc}{\mbox{$M_{\rm encl}$}}
\newcommand{\vlos}{\mbox{$v_{\rm los}$}}
\newcommand{\sigmalos}{\mbox{$\sigma_{\rm los}$}}
\newcommand{\sigmaloc}{\mbox{$\sigma_{\rm los,local}$}}
\newcommand{\avir}{\mbox{$\alpha_{\rm vir}$}}
\newcommand{\tff}{\mbox{$t_{\rm ff}$}}
\newcommand{\tdiss}{\mbox{$t_{\rm diss}$}}
\newcommand{\dvdl}{\mbox{${\rm d}\vlos/{\rm d}l$}}
\newcommand{\amin}{\mbox{${\rm arcmin}$}}
\newcommand{\asec}{\mbox{${\rm arcsec}$}}
\newcommand{\casa}{{\sc CASA}\xspace}
\newcommand{\be}{\begin{equation}}
\newcommand{\ee}{\end{equation}}
\newcommand{\bea}{\begin{eqnarray}}
\newcommand{\eea}{\end{eqnarray}}

\title{\vspace{-5mm}The dynamical evolution of molecular clouds near the Galactic Centre -- II. Spatial structure and kinematics of simulated clouds\vspace{-4mm}}

\author{J.~M.~D.~Kruijssen,$^{1,2}$\thanks{kruijssen@uni-heidelberg.de}
J.~E.~Dale,$^3$
S.~N.~Longmore,$^4$
D.~L.~Walker,$^{5,6}$
J.~D.~Henshaw,$^2$
\newauthor
S.~M.~R.~Jeffreson,$^1$
M.~A.~Petkova,$^1$
A.~Ginsburg,$^7$
A.~T.~Barnes,$^{4,8,9}$
C.~D.~Battersby,$^{10}$
\newauthor
K.~Immer,$^{11}$
J.~M.~Jackson,$^{12}$
E.~R.~Keto,$^{13}$
N.~Krieger,$^2$
E.~A.~C.~Mills,$^{14}$
\newauthor
\'A.~S\'anchez-Monge,$^{15}$
A.~Schmiedeke,$^8$
S.~T.~Suri$^{2,15}$ and
Q.~Zhang$^{13}$
\\
$^1$Astronomisches Rechen-Institut, Zentrum f\"{u}r Astronomie der Universit\"{a}t Heidelberg, M\"{o}nchhofstra\ss e 12-14, 69120 Heidelberg, Germany\\
$^2$Max-Planck Institut f\"{u}r Astronomie, K\"{o}nigstuhl 17, 69117 Heidelberg, Germany\\
$^3$Centre for Astrophysics Research, University of Hertfordshire, Hatfield, AL10 9AB, UK\\
$^4$Astrophysics Research Institute, Liverpool John Moores University, IC2, Liverpool Science Park, 146 Brownlow Hill,
Liverpool L3 5RF, UK\\
$^5$Joint ALMA Observatory, Alonso de Cordova 3107, Vitacura, Santiago, Chile\\
$^6$National Astronomical Observatory of Japan, Alonso de Cordova 3788, 61B Vitacura, Santiago, Chile\\
$^7$National Radio Astronomy Observatory, P.O. Box O 1009, Lopezville Drive, Socorro, NM 87801, USA\\
$^8$Max-Planck Institut f\"{u}r Extraterrestrische Physik, Giessenbachstra\ss e 1, 85748 Garching, Germany\\
$^9$Argelander Institute for Astronomy, University of Bonn, Auf dem H\"{u}gel 71, 53121 Bonn, Germany\\
$^{10}$University of Connecticut, Department of Physics, 2152 Hillside Road, U-3046, Storrs, CT 06269, USA\\
$^{11}$Joint Institute for VLBI ERIC (JIVE), Postbus 2, NL-7990 AA Dwingeloo, the Netherlands\\
$^{12}$School of Mathematical and Physical Sciences, University of Newcastle, University Drive, Callaghan NSW 2308, Australia\\
$^{13}$Harvard-Smithsonian Center for Astrophysics, 60 Garden Street, Cambridge, MA 02138, USA\\
$^{14}$Physics Department, Brandeis University, 415 South Street, Waltham, MA 02453, USA\\
$^{15}$I.~Physikalisches Institut, Universit\"{a}t zu K\"{o}ln, Z\"{u}lpicher Stra\ss e 77, 50937 K\"{o}ln, Germany
\vspace{-6mm}
}

\begin{document}

\date{Accepted 2019 February 5. Received 2019 February 1; in original form 2018 April 19.\vspace{-4mm}}

\pagerange{\pageref{firstpage}--\pageref{lastpage}} \pubyear{2019}

\maketitle

\label{firstpage}

\begin{abstract}
The evolution of molecular clouds in galactic centres is thought to differ from that in galactic discs due to a significant influence of the external gravitational potential. We present a set of numerical simulations of molecular clouds orbiting on the 100-pc stream of the Central Molecular Zone (the central $\sim500~\pc$ of the Galaxy) and characterise their morphological and kinematic evolution in response to the background potential and eccentric orbital motion. We find that the clouds are shaped by strong shear and torques, by tidal and geometric deformation, and by their passage through the orbital pericentre. Within our simulations, these mechanisms control cloud sizes, aspect ratios, position angles, filamentary structure, column densities, velocity dispersions, line-of-sight velocity gradients, spin angular momenta, and kinematic complexity. By comparing these predictions to observations of clouds on the Galactic Centre `dust ridge', we find that our simulations naturally reproduce a broad range of key observed morphological and kinematic features, which can be explained in terms of well-understood physical mechanisms. We argue that the accretion of gas clouds onto the central regions of galaxies, where the rotation curve turns over and the tidal field is fully compressive, is accompanied by transformative dynamical changes to the clouds, leading to collapse and star formation. This can generate an evolutionary progression of cloud collapse with a common starting point, which either marks the time of accretion onto the tidally-compressive region or of the most recent pericentre passage. Together, these processes may naturally produce the synchronised starbursts observed in numerous (extra)galactic nuclei.
\end{abstract}

\begin{keywords}
stars: formation -- ISM: clouds -- ISM: evolution -- ISM: kinematics and dynamics -- Galaxy: centre -- galaxies: ISM.\vspace{-6mm}
\end{keywords}

\section{Introduction} \label{sec:intro}

The evolution of molecular clouds and their ability to form stars is a fundamental process in astrophysics, shaping the stellar populations and galaxies that populate the Universe. Over the past decades, star formation within molecular clouds has been studied in extensive detail in the solar neighbourhood, the Galactic disc, and the Magellanic Clouds (see e.g.~\citealt{kennicutt12} for a review). However, when considered in a cosmological context, this range of environments is extremely limited and unrepresentative of the conditions under which most stars in the Universe formed. For instance, the gas densities, pressures, and velocity dispersions in these local-Universe environments \citep[e.g.][]{heyer15} are factors-of-several to orders of magnitude lower than at the time of the peak cosmic star formation history \citep[e.g.][]{swinbank12}.

Thanks to major Galactic plane surveys across a wide wavelength range \citep[e.g.][]{oka98,schuller09,schuller17,molinari10b,bally10,aguirre11,walsh11,purcell12,jones12,ginsburg13,jackson13,krieger17,longmore17}, as well as the arrival of the Atacama Large Millimeter/submillimeter Array (ALMA), it is now possible to study cloud evolution and star formation near the Galactic Centre, i.e.~in the Central Molecular Zone (CMZ, the central $\sim500~\pc$) of the Milky Way. This represents a major extension of the conditions probed traditionally and is crucial for a fundamental understanding of cloud evolution and star formation for the following reasons.
\begin{enumerate}
\item
The gas densities \citep[e.g.][]{bally87,rathborne15}, pressures \citep[e.g.][]{oka01,rathborne14b}, temperatures \citep[e.g.][]{huettemeister93,ao13,mills13,ginsburg16,krieger17}, and velocity dispersions \citep[e.g.][]{shetty12,henshaw16,kauffmann17} of CMZ clouds are similar to those in high-redshift galaxies at the time of the peak cosmic star formation rate \citep{kruijssen13c}. Insights drawn from star formation within CMZ clouds are thus likely to be more representative for how most stars in the Universe formed than those from solar neighbourhood studies.
\item
The evolution of CMZ clouds and their ability to form stars has been suggested to be closely coupled to galactic (and orbital) dynamics and strong shearing motions \citep{longmore13b,kruijssen14b,krumholz15,federrath16,meidt18b}. This close relation may give rise to episodic star formation \citep[e.g.][]{krumholz17} and places the current CMZ near a star formation minimum, plausibly explaining why the CMZ has an unusually low star formation efficiency \citep[also see \citealt{guesten83,taylor93,kauffmann13,kauffmann17,lu19}]{longmore13,barnes17}. A similar underproduction of stars per unit `dense' ($\rho\ga10^4~\cmc$) gas has recently been found in extragalactic centres \citep{usero15,bigiel16,gallagher18}, implying that the results found for the Galactic CMZ also extend to other galaxies. The relation between star formation activity and galactic dynamics may play an important role in setting the bottlenecks (or avenues) towards feeding supermassive black holes in galactic nuclei \citep[e.g.][]{kruijssen17} and may apply more generally to galactic discs \citep[e.g.][]{meidt18}. The highly dynamical environment of the CMZ provides the best opportunity for characterising the pertinent physics, especially thanks to its close proximity.
\item
A subset of the CMZ clouds, i.e.~the `dust ridge' (\citealt{lis94}, also see Section~\ref{sec:maps}), may follow an absolute evolutionary time sequence, potentially providing the unique opportunity of studying cloud evolution, star formation, and feedback as a function of absolute time \citep[e.g.][]{longmore13b}. A variety of recent papers have shown that the gas in the inner CMZ ($R<120~\pc$) is situated on a stream that follows an eccentric orbit \citep{molinari11,kruijssen15,henshaw16} and marks the transition from highly supervirial gas at larger radii to nearly virialised gas on the stream \citep[e.g.][]{kruijssen14b,walker15,henshaw16b,kauffmann17b}. Similar structures are found in extragalactic observations \citep[e.g.][]{peeples06} and numerical simulations of galactic centres \citep{emsellem15,sormani18}. The tidal field at the radii spanned by this `100-pc stream' is fully compressive due to the steep slope of the enclosed mass distribution \citep{kruijssen15}.\footnote{Appendix~A of \citet{kruijssen15} shows that the enclosed stellar mass $\menc\propto r^{2.2}$ for the galactocentric radii of interest, yielding a radial tidal force of $T_{rr}=-\partial/\partial r (G\menc/r^2)=-0.2G\menc/r^3<0$. The azimuthal and vertical tidal forces are more compressive by definition in axisymmetric potentials.} The strength of this tidal compression peaks when clouds pass through pericentre, which has led to the suggestion that the pericentre passage may nudge the clouds into gravitational collapse as they condense out of the supervirial, diffuse medium \citep{longmore13b,rathborne14,kruijssen15}. This idea is supported by observations showing trends of increasing gas temperatures \citep{ginsburg16,krieger17} and possibly star formation activity \citep{immer12b,schmiedeke16,barnes17,walker18,ginsburg18} after pericentre passage, as well as by theoretical results showing that epicyclic perturbations (i.e.~pericentre passages) may influence cloud evolution in the CMZ prior to gravitational free-fall taking over \citep{jeffreson18b}. If true, these results have the major implication that the position along the 100-pc stream is an indicator of evolutionary age. However, such evolutionary sequences along the 100-pc stream are not always monotonic \citep{kauffmann17b}, prompting the question whether deviations from monotonicity may reflect differences in initial conditions \citep{kruijssen17}, or if other events may induce cloud collapse and subsequent star formation.
\end{enumerate}

In this paper, we present hydrodynamical simulations of gas clouds on the best-fitting orbit of the 100-pc stream from \citet{kruijssen15}. There exists a rich literature on simulations of isolated clouds, but studies on the evolution of single clouds in external potentials are limited in number. We discuss in detail how the gravitational potential and the pericentre passage affect the morphology and kinematics of the simulated clouds and compare the results to observed CMZ clouds. This way, we characterise the interplay between orbital dynamics and cloud evolution in the CMZ and find that the arrival of gas onto the 100-pc stream and its subsequent pericentre passage mark a transformational event that reproduces several key features of observed CMZ clouds. In a companion paper \citep{dale19}, we present the simulations in detail, discuss the general properties of clouds orbiting in external potentials relative to several control experiments, and investigate how the external potential affects the star formation activity of the orbiting clouds.

\section{Numerical simulations of gas clouds on an eccentric orbit in the CMZ potential} \label{sec:sims}

\subsection{Summary of the numerical model} \label{sec:model}

We carry out simulations of massive gas clouds in the CMZ environment \citep{dale19} using the state-of-the-art smoothed particle hydrodynamics (SPH) code {\sc gandalf} \citep{hubber18}. The simulations considered here include self-gravity using an octal tree, as well as hydrodynamics and artificial viscosity \citep{monaghan97}, with the gas following a barytropic equation of state with a critical density of $2.5\times10^7~\cmc$ (assuming a mean molecular weight per particle of $\mu_{\rm p}=2.35$). Gas above a density of $2.5\times10^6~\cmc$ is converted into sink particles, implying that the effective equation of state is isothermal, with an adopted temperature of $T=65~{\rm K}$ to represent the high temperature of CMZ gas \citep[e.g.][]{ao13,ginsburg16,krieger17}. At the adopted numerical resolution (see below), these sink particles do not represent individual stars, but stellar subclusters. We assign no further physical meaning to these particles, because we are interested in following the large-scale morphology and kinematics of the clouds. The sink particles are only included to prevent the simulations from reaching excessively high (and computationally costly) densities.

The presented simulations do not include stellar feedback or magnetic fields, because we aim to study the pure gravo-hydrodynamics of gas clouds on an eccentric orbit around the Galactic Centre. These mechanisms would act to slow down or truncate star formation. As a result, the simulated clouds will always collapse to form stars at an efficiency much higher than observed or obtained by simulations that do include feedback and/or magnetic fields. While the present paper does not report on any of the star formation characteristics of the simulations, we quantify the relative influence of orbital dynamics on the star formation efficiency in \citet{dale19}.

The full suite of simulations presented in \citet{dale19} contains 36 models of $10^6$ SPH particles each, consisting of 12 different sets of initial conditions followed in three different tidal environments. Here we consider a total of 5 different simulations that represent the subset of `tidally-virialised' clouds from \citet[also see below]{dale19} and are moving on the \citet{kruijssen15} orbit in the gravitational potential generated by the mass distribution from \citet{launhardt02}, with a vertical flattening parameter $q_\Phi=0.63$ (see Appendix~A of \citealt{kruijssen15}). For comparison, we also consider five `control runs', in which the clouds are moving on a circular orbit in the same background potential. These simulations follow the evolution of spherical clouds with a Gaussian initial density profile, which we truncate at the radius $R_{\rm t}$, where the local volume density has dropped to 5~per~cent of the central value (corresponding to 2.45 standard deviations or 1.70 times the half-mass radius $R_{\rm h}$ of the truncated Gaussian). The clouds have a divergence-free initial turbulent velocity field, i.e.~without any net compression or expansion. We have carried out several random realisations of the velocity field and select the one with a net angular velocity opposite to the orbital motion of $1~\myr^{-1}$. This way, the spin angular momentum and angular velocity are roughly consistent with what a cloud would have if it were given an infinitesimal perturbation to induce its contraction towards its centre of mass within the shearing interstellar medium in the CMZ potential (cf.~its shear-driven angular velocity of $r{\rm d}\Omega/{\rm d}r\approx-0.7~\myr^{-1}$, caused by the fact that the circular velocity increases sub-linearly with galactocentric radius). This is also consistent with the regular spacing of clouds observed on the 100-pc stream \citep{henshaw16b}.

\begin{table}
 \centering
  \begin{minipage}{77mm}
\caption{Initial conditions of the simulations.}
\label{tab:ics}
\begin{tabular} {@{}lccccccc@{}}
  \hline
 Simulation & $M$ & $R_{\rm t}$ & $\sigma$ & $\avir$ & $\rho$ & $\Sigma$ & $\tff$ \\ 
  \hline
  fiducial & 7.7 & 13.6 & 24.1 & 9.4 & 1.3 & 1.3 & 0.94  \\ 
  low density & 13.4 & 23.5 & 41.2 & 27.4 & 0.4 & 0.8 & 1.63 \\ 
  high density & 4.5 & 7.8 & 13.9 & 3.1 & 3.8 & 2.3 & 0.54 \\ 
  low vel.~disp. & 1.0 & 6.8 & 12.1 & 9.4 & 1.3 & 0.7 & 0.94 \\ 
  high vel.~disp. & 26.1 & 20.4 & 36.2 & 9.4 & 1.3 & 2.0 & 0.94 \\ 
  \hline
\end{tabular}
Note: All listed quantities are evaluated over the full volume of the clouds, i.e.~out to the truncation radius $R_{\rm t}$, which is about 1.7 times the half-mass radius. Units: $M$ in $10^5~\msun$, $R_{\rm t}$ in $\pc$, $\sigma$ in $\kms$, $\rho$ in $10^3~\cmc$, $\Sigma$ in $10^3~\msun~\pc^2$, and $\tff$ in $\myr$.
\end{minipage}
\end{table}
The global properties of the clouds are uniquely set by the mean volume density $\rho$, virial parameter $\avir\equiv 2T/|V|$ (with $T$ the kinetic energy and $V$ the gravitational potential energy),\footnote{For this definition, virial equilibrium is achieved at $\avir=1$.} and velocity dispersion $\sigma$, implying that the masses and radii vary. We first generate clouds with $\rho=\{0.4, 1.3, 3.8\}\times10^3~\cmc$, $\avir=\{0.87, 2.6, 7.8\}$, and $\sigma=\{6.3, 12.7, 19.0\}~\kms$, with the intermediate values representing the fiducial model (these choices are justified below). Relative to the fiducial model, we obtain six non-fiducial models by changing one parameter at a time. All non-fiducial models are derived from the fiducial model by scaling the particle masses, as well as their position and velocity vectors, such that the resulting set of initial conditions is homologous. We then isotropically adjust the velocity dispersions such that the internal turbulent energy balances the compressive tidal energy due to the background potential. Without such a `tidal virialisation' (as in the `self-virialised' models of \citealt{dale19}) by elevating the turbulent velocity dispersion and the corresponding virial ratio, the clouds undergo rapid gravitational collapse. This tidal virialisation fixes the virial parameter, because it becomes inversely related to the cloud density, thus leaving us with five models in total. The resulting initial velocity dispersions and virial parameters of the clouds span $\sigma=12.1$--$41.2~\kms$ and $\avir=3.1$--$27.4$. All parameters describing the initial conditions are summarised in \autoref{tab:ics}.

The initial column and volume densities are chosen to be similar to those observed in the CMZ upstream of the dust ridge (at $-0\fdg7<l<0\fdg0$ and $-0\fdg1<b<0\fdg1$), which are thought to have recently condensed out of the diffuse medium \citep{henshaw16b}, spanning $\Sigma=0.5$--$1.5\times10^3~\msun~\pc^{-2}$ and $\rho=2$--$7\times10^3~\cmc$, respectively \citep{henshaw17}. While these densities are already similar to those used in our initial conditions, it is important to note that these observations probe projected radii smaller by a factor of $\sim7$. Using the observed size-linewidth relation $\sigma\propto R^{0.7}$ \citep[e.g.][]{shetty12,kauffmann17} and assuming virial equilibrium ($M\propto R\sigma^2$), we expect that $\Sigma\propto R^{0.4}$ and $\rho\propto R^{-0.6}$. Extrapolating the densities of all clouds from \citet{henshaw17} to the spatial scales of the simulated clouds, we obtain $\Sigma=0.8$--$3.3\times10^3~\msun~\pc^{-2}$ and $\rho=0.6$--$3.2\times10^3~\cmc$, in excellent agreement with \autoref{tab:ics}. Similarly, the velocity dispersions are consistent with the observed size-linewidth relation \citep[e.g.][]{shetty12,kruijssen13c,kauffmann17}, which at the radii listed in \autoref{tab:ics} require $\sigma=5$--$40~\kms$. Again, this matches the range spanned by our set of initial conditions, indicating that the simulations capture the full range of physical conditions observed upstream of the dust ridge.

\begin{table}
 \centering
  \begin{minipage}{64mm}
\caption{Simulations, snapshots, and corresponding orbital positions in the \citet{kruijssen15} model used for generating \autoref{fig:map}.}
\label{tab:snaps}
\begin{tabular} {@{}lcccc@{}}
  \hline
 Cloud & Simulation & $t-t_0$ & $l_{\rm orb}$ & $b_{\rm orb}$ \\
 & & [$\myr$] & [$\degr$] & [$\degr$] \\ 
  \hline
  Pebble & low density & 0.58 & 0.04 & 0.03 \\ 
  Brick & high density & 0.74 & 0.25 & 0.02 \\ 
  Clouds c/d & low vel.~disp. & 0.84 & 0.37 & 0.01 \\ 
  Clouds e/f & fiducial & 0.87 & 0.41 & 0.01 \\ 
  Sgr B2 & high vel.~disp. & 1.14 & 0.64 & $-$0.05 \\ 
  Sgr B2+ & fiducial & 1.27 & 0.71 & $-$0.07 \\ 
  \hline
\end{tabular}
Note: The simulated clouds' centres of mass may deviate from the above orbital $\{l,b\}$ coordinates by up to $0\fdg06$, depending on their spatial extent and fragmentation. The time $t_0$ indicates the time at which the simulations start, such that $t-t_0$ represents the time since the beginning of the simulations.
\end{minipage}
\end{table}

\begin{figure*}
\includegraphics[width=0.98\hsize]{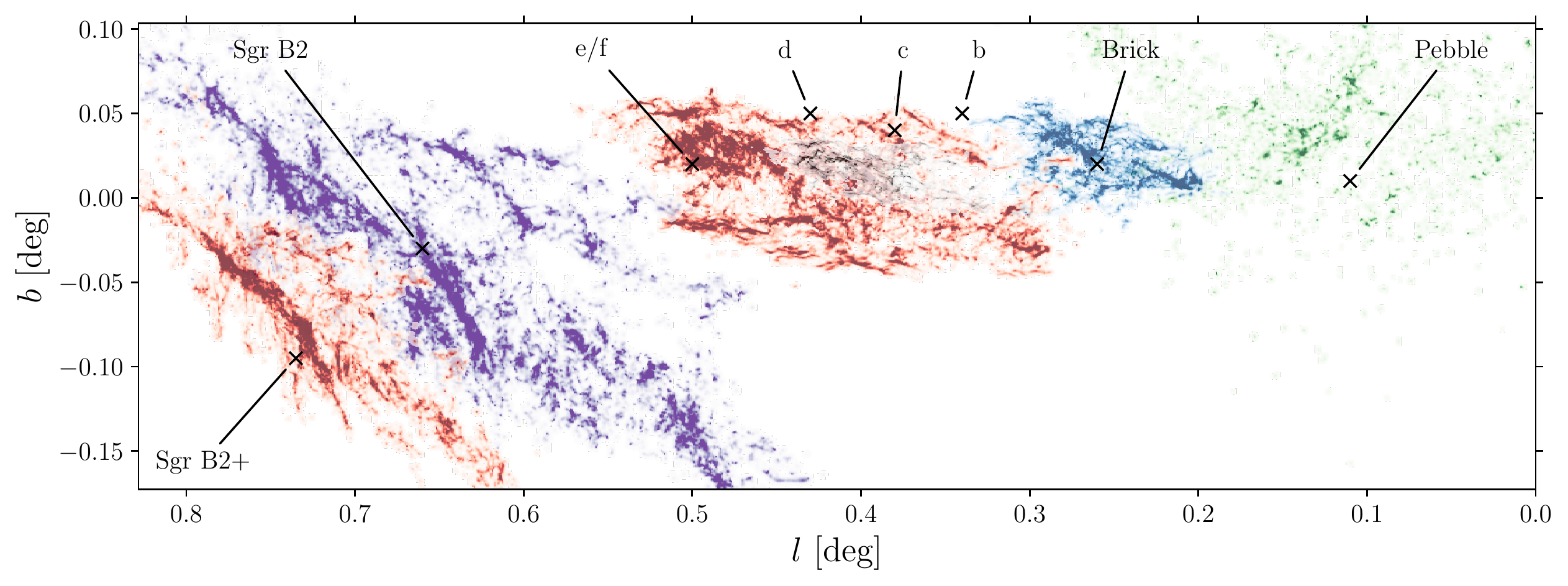}%
\vspace{-2mm}\caption{
\label{fig:colours}
Illustration of how the simulations in the $\{l,b\}$ plane at several different snapshots are combined to represent the observed `dust ridge' of the Galactic CMZ (see the text). The colours highlight how this map is constructed from the different simulations listed in \autoref{tab:ics}, using the snapshots listed in \autoref{tab:snaps}. Specifically, \{red, green, blue, grey, purple\} indicates the \{fiducial, low density, high density, low velocity dispersion, high velocity dispersion\} simulation. The resulting composite column density maps are shown in \autoref{fig:map}. For reference, the annotations show the locations of various {\it observed} dust ridge clouds.
}
\end{figure*}

We deliberately model clouds with masses higher than those of most of the observed CMZ clouds, so that we can model the condensation of such clouds out of a more extended gas reservoir \citep[cf.][]{henshaw16b}. We initialise the clouds shortly after the apocentre passage preceding the dust ridge in the \citet{kruijssen15} model, at coordinates $\{l,b\}=\{-0\fdg65,-0\fdg07\}$ and a radius of $r_0=90~\pc$ \citep[corresponding to $t_0=-2.46~\myr$ in Appendix~C of][]{kruijssen15}. Each simulation then follows a single cloud through pericentre, across the dust ridge, past the position of Sgr~B2. This way, the simulations capture the evolution of the clouds across a complete radial orbital oscillation, from their diffuse state near apocentre through the dynamical perturbations induced by their orbit in the gravitational potential. Throughout the paper, we use a non-rotating coordinate system centred on the origin of the Galactic coordinate system, i.e.~on $\{l,b\}=\{0\fdg0,0\fdg0\}$, with the $x$-axis pointing towards negative Galactic longitudes, the $y$-axis pointing from the observer to the Galactic centre (with Sgr~A$^*$ at $y=0$), and the $z$-axis pointing towards positive Galactic latitudes.

\subsection{Column density maps} \label{sec:maps}

In order to perform a first, visual comparison of the simulated clouds to the observed CMZ clouds on the dust ridge, we combine a number of snapshots from the different simulations, taken at different times to match the positions of the observed clouds, and generate column density maps of the gas with densities $\rho\geq10^4~\cmc$. This density threshold is chosen to suppress the (initially) extended cloud envelopes in the simulations and represent the high-density gas found in the CMZ, as traced by the bulk of the dust emission \citep[e.g.][]{longmore13} and molecules such as HCN and NH$_3$ \citep[e.g.][]{mills13,rathborne14,krieger17}. We consider three projections of the simulation snapshots, i.e. top-down $\{l,y\}$, plane-of-sky $\{l,b\}$, and position-velocity $\{l,\vlos\}$.\footnote{Throughout this paper, we adopt a distance to the Galactic Centre of $d=8.3~\kpc$ \citep{reid14} when converting between angular and physical sizes, for which $\{1\degr,1\arcmin,1\arcsec\}\approx\{145,2.41,0.040\}~\pc$.}

The snapshots are selected to visually best represent the real CMZ clouds in the $\{l,b\}$ plane in terms of their spatial extents, column densities, and orientations. This is a necessary step, because there are no constraints on the initial conditions of the clouds. The relevant question to ask is thus whether any of them reproduce the $\{l,b,\vlos\}$ structure of the observed dust ridge clouds. Even though this optimises the agreement between simulations and observations, it is an interesting comparison to make, because the initial conditions of the simulations span the full range of physical conditions seen in the progenitors of the dust ridge clouds. As we will see below, the simulated evolution of this variety of progenitors reproduces the observed variety of dust ridge clouds, which is an important test of the model's self-consistency. This enables the interpretation of several key observables in terms of the underlying physical processes. In doing so, we caution that this interpretation should be carried out in terms of global characteristics or systematic trends only. Direct comparisons between any observed and simulated cloud should be made with caution, as the detailed structure (e.g.~individual filamentary structures and positions of protostellar cores) is determined entirely by the specific realisation of the initial conditions.

\begin{figure*}
\includegraphics[width=0.97\hsize]{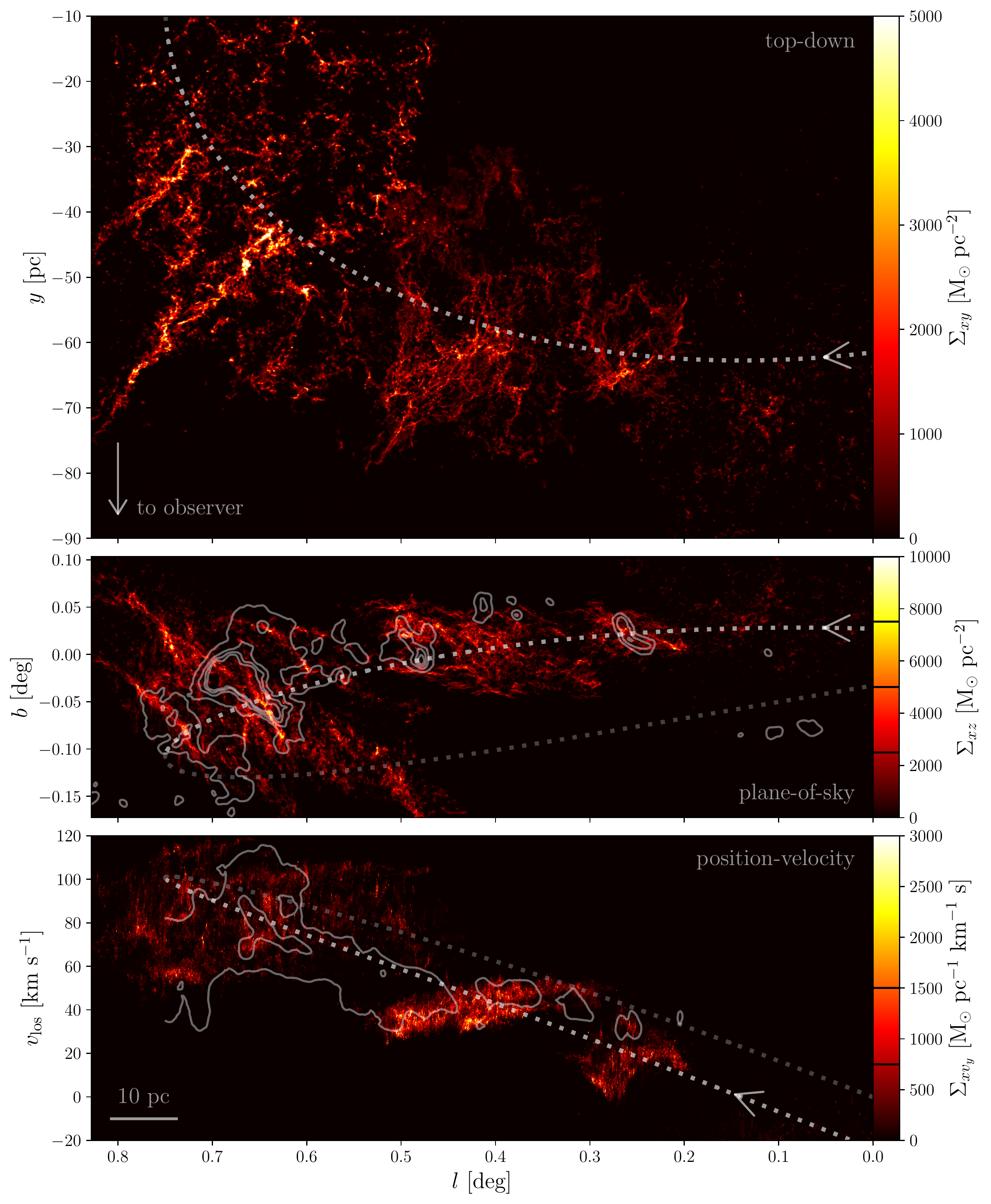}%
\vspace{-2mm}\caption{
\label{fig:map}
Combined column density maps of the simulations at several different snapshots, chosen to best represent the observed clouds on the CMZ `dust ridge' (see the text). The orbital solution of \citet{kruijssen15} is shown as the light grey line, with the segment thereof covered by the simulations shown in a brighter shade and arrows indicating the direction of motion. Top panel: top-down projection. Middle panel: plane-of-sky projection. For comparison, the observed column density map of the gas traced by cold dust (from HiGAL, see Battersby et al.~in prep.) is shown by contours at $\Sigma_{xz}=\{2.5,5,7.5,10\}\times10^3~\msun~\pc^{-2}$ (black lines in the colour bar). Bottom panel: position-velocity projection. This figure shows that the main morphological and kinematic features of the observed dust ridge clouds can be reproduced by drawing a population of clouds from the initial conditions of \autoref{tab:ics} (see \autoref{tab:snaps}) and simulating their dynamical evolution in the gravitational potential of the CMZ. For comparison, the observed position-velocity distribution of the molecular gas in the dust ridge traced by $^{13}$CO(2-1) \citep{ginsburg16} is shown by contours at $\Sigma_{xv_y}=\{0.75,1.5\}\times10^3~\msun~\pc^{-1}~{\rm km}^{-1}~{\rm s}$ (black lines in the colour bar). An animated version of this figure is available as online Supporting Information.
}
\end{figure*}

The adopted snapshots are summarised in Table~\ref{tab:snaps} and their relative positioning in the $\{l,b\}$ plane is shown in \autoref{fig:colours}. These clouds represent the main ingredients of the dust ridge (the Brick or G0.253+0.016, Clouds c/d/e/f, and Sgr~B2), and also include a low-density cloud upstream of the Brick (which we dub the `Pebble', at $\{l,b\}=\{0\fdg11,0\fdg00\}$) and the downstream continuation of the Sgr~B2 complex (`Sgr~B2+', at $\{l,b\}=\{0\fdg72,-0\fdg08\}$). Three different projections of the combined maps are shown in \autoref{fig:map}. An animated version of this figure is available as online Supporting Information. For comparison, \autoref{fig:map} also includes contours showing the observed gas column densities derived from the cold dust (HiGAL, Battersby et al.~in prep.) and contours showing the observed position velocity structure in $^{13}$CO(2-1) \citep{ginsburg16}.\footnote{Throughout this work, we transform velocities from the local standard of rest to the rest frame of Sgr~A* by adding $U_\odot=14~\kms$ \citep{schoenrich12}, consistently with \citet{kruijssen15}.} These have considerably lower spatial resolutions ($25\arcsec$ and $30\arcsec$, respectively, corresponding to $1.0~\pc$ and $1.2~\pc$) than the simulations ($\sim0.1~\pc$), which leads to trivial differences in spatial structure between the simulations and observations. None the less, these maps provide a good point of reference, because they can be converted into physical units with relatively few assumptions.\footnote{ We adopt a mean molecular weight per ${\rm H}_2$ molecule of $\mu_{{\rm H}_2}=2.8$ to convert dust column densities to mass densities. To translate the $^{13}$CO flux to a mass density, we assume a $^{13}$CO-to-H$_2$ conversion factor of $\alpha({\rm ^{13}CO})=22.8~\msun~\pc^{-2}~({\rm K}~\kms)^{-1}$ \citep{schuller17,cormier18}, which is particularly appropriate in high-column density environments like the CMZ.} A number of qualitative properties of the simulated clouds immediately stand out. We list these here and quantify several of them further in Section~\ref{sec:quant} below.

\begin{enumerate}
\item
{\it Fragmentation and line-of-sight extension}: The top panel of \autoref{fig:map} shows that the clouds collapse and fragment as the turbulent energy dissipates. At the same time, the external gravitational potential causes the fragments within each cloud to roughly maintain their mutual separation, thus inhibiting global collapse across the full $2R_{\rm t}=15$--$50~\pc$ size scales modelled here. The most massive of these fragments match the masses of young massive cluster progenitors observed on the dust ridge \citep[see e.g.][]{walker15} and proceed to collapse. The extensions towards the observer at high longitudes result from the shear-driven dispersal of the outer layers of the clouds, which may explain observational hints of expanding outer layers in the Brick \citep{rathborne14,bally14}. These extensions are a common feature across most of our simulations after a few cloud dynamical times, which strongly suggests that the real CMZ clouds at positive Galactic longitudes may have significant depth along the line of sight, spanning a typical length scale comparable to their major axis in the plane of the sky. Similar, radially extended `feathers' are prevalent in dust extinction maps of gas clouds in extragalactic centres \citep[e.g.][]{peeples06}. Our simulations suggest that such feathers arise due to shearing motions in the galactic potential. When projected along the line of sight, these feathers could plausibly be (mis)interpreted as e.g.~the base of a spiral arm or a cloud-cloud collision due to line-of-sight confusion or multiple velocity components.
\item
{\it Vertical compression, large-scale filamentary structure, and inclination}: The middle panel of \autoref{fig:map} shows that the clouds attain a vertically compressed morphology just after they have passed through pericentre (which occurs at $l=-0\fdg16$). This is mainly caused by the tidal field, which is considerably more compressive in the vertical direction than azimuthally (see Section~\ref{sec:struc}). The compression generates layers and leaves the clouds highly substructured, with flocculent filamentary features running in the longitudinal direction, much akin to the structure of the Brick as observed with ALMA \citep{rathborne15} and of clouds b, d, and e as observed with the Submillimeter Array \citep[SMA,][]{walker18}. The tilt of the clouds towards positive longitudes and latitudes arises due to a torque experienced by the clouds as they move through pericentre above the Galactic plane. On its approach to pericentre passage, the leading end of the cloud arrives first and is pulled upwards. This combination of inclination and flattening is a common feature across all of our simulations and matches that of the observed contours at the positions of the Brick, clouds e/f, Sgr~B2, and Sgr~B2+.
\item
{\it Velocity gradient, shear, and kinematic complexity}: The bottom panel shows that the clouds develop a velocity gradient with a direction opposite to the orbital motion. This counter-gradient is driven by shear, which causes the side of the cloud facing the Galactic centre to move faster than the side facing away from the Galactic centre (see Section~\ref{sec:kin}). Again, this is a common feature of all models and qualitatively matches the velocity gradient of the Brick, clouds e/f, and the Sgr~B2 complex. However, it is also quite sensitive to the realisation of the initial velocity field, which was chosen to be consistent with the onset of contraction from a shearing gas reservoir (see Section~\ref{sec:model}). Clouds with zero initial spin angular momentum exhibit significantly weaker velocity gradients. Finally, this panel also reveals significant kinematic complexity in the Sgr~B2 region, which is caused by the superposition of clouds along the line of sight as the orbit curves off, away from the observer. This matches the observed, complex kinematic structure of the cloud complex spanned by Sgr~B2 and Sgr~B2+ \citep[e.g.][]{henshaw16} and provides a plausible alternative to the currently canonical interpretation of this complexity as a key piece of evidence for a cloud-cloud collision \citep[e.g.][]{mehringer93,hasegawa94,sato00,jones08}. We find that no such collision is necessary to generate the large degree of kinematic complexity in the Sgr~B2 region. Instead, this is naturally reproduced by the combination of fragmentation and the orbital geometry.
\end{enumerate}

In summary, the clouds' evolution in a background potential and the pericentre passage initially turns them into inclined, sub-structured, spinning pancakes, before the combination of fragmentation and orbital curvature generates highly complex $\{l, b, \vlos\}$ structures. We find that a wide range of properties of the observed dust ridge clouds (contours in \autoref{fig:map}) are reproduced by making a suitable selection from the set of simulations. As stated before, this is an interesting result, because the initial conditions of the simulations were chosen to be representative of the observed range of physical conditions (i.e.~column densities, volume densities, and velocity dispersions) in the clouds upstream from pericentre (see Section~\ref{sec:model}). This shows that the morphology and kinematics of the dust ridge clouds are consistent with their deformation due to the background potential and the preceding pericentre passage as modelled here, under the plausible condition that their initial properties were similar to those currently observed upstream.

More broadly, \autoref{fig:map} illustrates how the evolution of molecular clouds in the CMZ is closely coupled to their orbital dynamics. As will be quantified further in the next sections, these results can be generalised to any (extra)galactic centre where the rotation curve turns over from being flat to $v\propto r^\beta$ with $1/2<\beta<1$, which is far from unusual \citep[see e.g.][]{miyamoto75,rubin80,persic95,errozferrer16}. For any rotation curve with such a turnover, the inflow of gas from larger galactocentric radii is predicted to stall due to a decrease of the shear, which leads to the accumulation of the gas in a ring-like structure like the 100-pc stream \citep{krumholz15}. During this accumulation, the orbital motion of the gas is synchronised by hydrodynamic forces, after which the stream fragments into clouds undergoing synchronised, semi-ballistic motion on an eccentric orbit that retains some memory of the non-zero radial inflow velocity \citep[this behaviour has also been found in three-dimensional simulations, see e.g.][]{emsellem15}.\footnote{The eccentric stream thus exists within the ring-like region of minimal (but non-zero) shear, which is $45<r/\pc<115$ in the gravitational potential of \citet[see Section~\ref{sec:kin} below and \citealt{krumholz15}]{launhardt02} and closely matches the pericentre and apocentre radii of the stream ($r_{\rm p}=60~\pc$ and $r_{\rm a}=120~\pc$, respectively).}

Our simulations show the clouds on the stream are torqued, compressed, and stretched, each of which has the potential to fundamentally change their subsequent evolution. Given the strength of these interactions, the dynamical coupling with the host galaxy may induce collapse and star formation in certain `hotspots', which appear as evolutionary streamlines of progressively advanced states of star formation and feedback. Such evolutionary sequences may be ubiquitous in extragalactic centres too, but they are unlikely to be strictly monotonic -- \autoref{tab:snaps} and \autoref{fig:map} show that some diversity of initial conditions is needed to reproduce the observed CMZ clouds, indicating that the responses of these clouds to the background potential and pericentre passage may differ. In the next section, we quantify how the morphology and kinematics of the clouds are affected by their orbital dynamics.

\section{Quantitative properties of simulated clouds} \label{sec:quant}

We now turn to a quantitative discussion of the simulations, focussing on the spatial structure and kinematics of the clouds. In addition to an extensive discussion of the simulations following clouds on eccentric orbits, we also include a brief comparison to a set of control simulations of clouds on circular orbits. The full set of clouds on circular orbits is presented in Appendix~\ref{sec:circ}.

\subsection{Spatial structure} \label{sec:struc}

The morphological evolution of the clouds is quantified in \autoref{fig:struc} for all five simulations. For reference, a colour composite of the inner CMZ with the orbital model highlighted is included as the top panel. The figure shows several key (observable) properties of the simulated clouds as a function of Galactic longitude, i.e.~the dimensions of the clouds in the plane of the sky, their aspect ratios, and their plane-of-sky column densities. All quantities are calculated using the gas with densities $\rho\geq10^4~\cmc$ as in \autoref{fig:map}, which represents the centrally concentrated components of the clouds and is traced by most of the dust emission \citep{longmore13}, as well as molecular line tracers \citep[such as HCN and NH$_3$, see e.g.][]{mills13,rathborne14}. The dimensions of the clouds in the Galactic longitudinal ($\delta x$) and latitudinal ($\delta z$) directions are obtained by calculating the centre of mass in each coordinate ($x_{\rm cm}$ or $z_{\rm cm}$) and finding the corresponding intervals that enclose half of the mass (i.e.~$x_{\rm cm}\pm\delta x$ or $z_{\rm cm}\pm\delta z$). The aspect ratios follow as $\delta z/\delta x$.\footnote{Note that we use Cartesian coordinates $\{x,z\}$ to denote positions within a given simulation snapshot, whereas $\{l,b\}$ refer to position along the orbit and thus include the motion and time evolution of the clouds.} Finally, the average column densities of the clouds ($\Sigma$) are calculated in a mass-weighted fashion over the rectangle spanned by a cloud's half-mass intervals (thus together enclosing less than half of its mass) as
\be
\label{eq:column}
\Sigma = \frac{\int_{x_{\rm cm}-\delta x}^{x_{\rm cm}+\delta x}\int_{z_{\rm cm}-\delta z}^{z_{\rm cm}+\delta z} \Sigma_{\rm local}(x,z)^2 {\rm d}x{\rm d}z}{\int_{x_{\rm cm}-\delta x}^{x_{\rm cm}+\delta x}\int_{z_{\rm cm}-\delta z}^{z_{\rm cm}+\delta z} \Sigma_{\rm local}(x,z) {\rm d}x{\rm d}z} ,
\ee
with $\Sigma_{\rm local}(x,z)$ the local column density at the coordinate $\{x,z\}$. The weighting by mass emphasises the column density at which most of the mass resides and makes it insensitive to the total spatial extent of the clouds. This is desirable, because the simulations follow the condensation of clouds from a gas reservoir that is initially larger than the observed clouds (see \autoref{tab:ics} and Section~\ref{sec:model}). In addition, we prevent dilution by focussing a window on the cloud centre of mass. By taking a mass-weighted average over this region of interest rather than an area-weighted one, we predict the typical column density expected to be observed in CMZ clouds.

\subsubsection{Cloud dimensions and aspect ratios}

\begin{figure}
\includegraphics[width=\hsize]{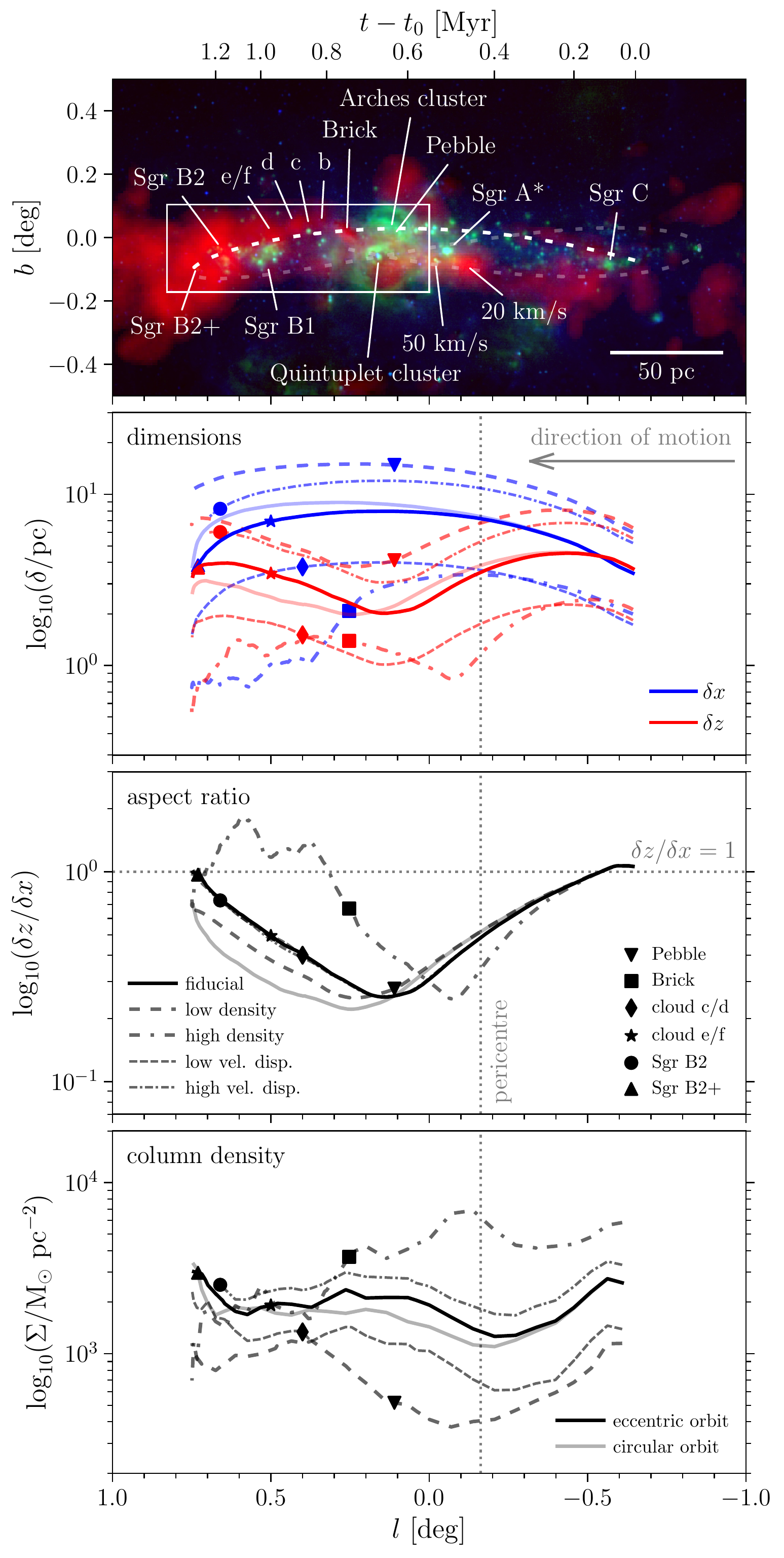}%
\vspace{-2mm}\caption{
\label{fig:struc}
Morphological evolution of the five simulated clouds. Panel~1: three-colour composite reference image of the CMZ, with HOPS NH$_3(1,1)$ in red \citep{walsh11,purcell12}, MSX 21.3~$\mu$m in green \citep{egan98,price01}, and MSX 8.3~$\mu$m in blue. The orbital model is shown as the dotted line, with the opaque white part indicating the trajectory covered by the simulated clouds. The top axis indicates the time along this section of the orbit. The area shown in the middle panel in \autoref{fig:map} is shown as a white box. The labels indicate several key objects in the CMZ. Panel~2: evolution of the cloud dimensions, represented by their longitudinal ($\delta x$, blue) and latitudinal ($\delta z$, red) half-mass radii. Panel~3: evolution of the cloud aspect ratios, i.e. $\delta z/\delta x$. Panel~4: evolution of the mass-weighted cloud column densities. In all panels, the lines encode the different simulations and filled symbols mark the values predicted at the observed longitudes of several dust ridge clouds (obtained by connecting each of them to a simulation as in \autoref{tab:snaps} and \autoref{fig:map}), as indicated by the legends in panel~3. To quantify the effect of the non-zero orbital eccentricity, transparent lines show the fiducial cloud on a circular orbit for comparison. The vertical dotted line marks the position of pericentre. This figure quantifies how the morphological evolution of the clouds is shaped by their orbital motion through the gravitational potential of the CMZ.
}
\end{figure}

Focusing on the cloud dimensions and aspect ratios (second and third panels of \autoref{fig:struc}), we see that they all exhibit a very similar evolution. In the Galactic longitudinal direction, the clouds are stretched by a factor of 2 as they pass through pericentre. This is a combined result of the clouds' orbital rotation and the fact that shear extends them further in the azimuthal direction than the radial direction. The vertical extent of the clouds is affected differently -- in the Galactic latitudinal direction, the clouds are flattened by a factor of 0.5. This arises due to a combination of two effects.

Firstly, geometric convergence of the orbital structure towards pericentre implies that all dimensions perpendicular to the direction of motion are compressed by a factor of $r_{\rm p}/r_0\approx0.67$, with $r_0=90~\pc$ the initial radius and $r_{\rm p}=60~\pc$ the pericentric radius. Radially, this occurs because all orbits of individual mass elements within a cloud are compressed into a smaller radial interval as the cloud approaches pericentre. In the vertical direction, pericentre represents the convergence point for mass elements that are vertically offset initially. In itself, this factor of $r_{\rm p}/r_0$ is capable of explaining the behaviour found in the simulations. It naturally leads to extreme aspect ratios (combining differential acceleration and geometric convergence yields $v_{l,0}r_{\rm p}/v_{l,{\rm p}}r_0\approx0.39$) as found in the second panel, which reach latitudinal-to-longitudinal size ratios as low as $\delta z/\delta x=0.25$.

A second, possibly more important mechanism leading to a vertical compression is provided by the tidal field. Assuming a power law enclosed mass profile $\menc\propto r^\alpha$ and hence a rotation curve $v\propto r^{(\alpha-1)/2}$ with angular velocity $\Omega=v/r\propto r^{(\alpha-3)/2}$, the tidal tensor in the CMZ is given by:
\be
\label{eq:tides}
T_{ij} = \frac{\partial^2\Phi}{\partial x_i\partial x_j} =
 \begin{bmatrix}
  (2-\alpha)\Omega^2 & 0 & 0 \\
  0 & -\Omega^2 & 0 \\
  0 & 0 & -q_\Phi^{-2}\Omega^2
 \end{bmatrix} ,
\ee
where $\Phi$ is the gravitational potential, $x_i$ is the $i$th component of the position vector, $q_\Phi$ is the vertical-to-planar axis ratio of the background potential, which indicates its degree of flattening, and we have chosen the coordinate system such that the tensor follows the order of the radial ($r$), tangential ($\phi$), and vertical ($z$) directions.

For the eccentric orbit under consideration, with $q_\Phi=0.63$ and $\alpha=2.2$ \citep{kruijssen15}, all terms on the diagonal of the tidal tensor are negative and the tidal field is fully compressive. A fully compressive tidal field exists where $\alpha>2$. In the CMZ potential derived by \citet{launhardt02} that we use here, this corresponds to the radial range $45<r/\pc<115$ \citep{kruijssen15}. This nearly exactly matches the radial extent of the gas stream,\footnote{This may not be a coincidence. In the galactic centre models by \citet{krumholz15} and \citet{krumholz17}, the inward transport of gas is driven by shear, which exists for $\alpha<3$. The closer the rotation curve is to solid-body ($\alpha=3$), the slower the radial transport. In the CMZ, shear is minimal at a peak value of $\alpha\approx2.3$, which is reached in the compressive region $45<r/\pc<115$. As a result, the inflow slows down and gas accumulates, thus forming the observed gas stream, where it is capable of reaching high densities and becomes prone to gravitational collapse and star formation. Observations of the CMZ match this picture, as the gas on the 100-pc stream is only marginally stable, whereas the gas at larger radii is characterised by a \citet{toomre64} parameter $Q>5$ \citep{kruijssen14b}.} giving rise to the somewhat unusual situation in which there is non-zero shear, but the tidal radii of the clouds are elevated relative to those in an extensive tidal field. The radial, azimuthal, and vertical components of the tidal field have relative strengths $(T_{rr}:T_{\phi\phi}:T_{zz})=(0.2:1:2.6)$, indicating that the strongest compression takes place in the vertical direction. The compressive tidal field in the azimuthal direction partially cancels the orbital stretching by shear described above, which explains why the vertical extent of the clouds gets compressed more steeply than their longitudinal extent increases. Put simply, the evolution in a background potential turns the clouds into pancakes.

Even on a circular orbit, the geometry of the tidal field causes the clouds to be strongly flattened \citep[grey solid line in \autoref{fig:struc}, also see][]{dale19}, with an aspect ratio $\delta z/\delta x=1/2.6\approx0.38$. This compression is enhanced by shear-driven azimuthal extension of the clouds. In addition, each of the components of the tidal tensor increases in strength by a factor of $(r_{\rm p}/r_0)^{\alpha-3}\approx1.4$ as the cloud moves from apocentre to pericentre. This amplifies the tidal compression and causes clouds on eccentric orbits to be flattened more quickly (by up to $0.2~\myr$) than those on circular orbits. In the simulations, the tidal field thus sets the minimum aspect ratio of $\delta z/\delta x=0.25$, whereas the pericentre passage accelerates the time by which this is achieved. Note that the Galactic longitude at which the pericentre passage takes place generally does not mark maxima or minima in any of the observables. The clouds respond to external perturbations on a crossing time, which means that they `overshoot' the time of pericentre passage in terms of their structural (and kinematic, see Section~\ref{sec:kin}) evolution.

Because the dimensions of the clouds are directly observable, we do not normalise their size evolution by the initial dimensions in the second panel of \autoref{fig:struc}. However, it is clear that the relative size evolution exhibited by the range of modelled clouds is nearly identical, independently of the initial conditions (e.g.~density, velocity dispersion, or virial state). The high-density simulation represents the only exception, in that it evolves much more quickly than the other clouds, reaching its most extreme aspect ratio some $0\fdg2$ in longitude before the other simulations do. The more rapid evolution seen in this simulation is a simple result of its high density and short dynamical time -- the cloud initiates its collapse prior to the pericentre passage, dynamically decouples from the background potential and, as a result, amplifies the initial perturbation imposed by its orbital motion on a shorter time-scale.\footnote{The same dynamical decoupling due to a local dominance of self-gravity takes place in the other clouds on the scales of individual, dense fragments. The high-density simulation is the only model in which the global density is sufficiently high for the entire cloud to achieve this.} Given that the presented simulations are drawn from a representative set of initial conditions, the range spanned by the simulations provides a prediction for the dimensions expected to be observed for real CMZ clouds.

After the minimum aspect ratio has been reached, $\delta z/\delta x$ increases again, because the azimuthal extension of the clouds is projected along the line of sight as the orbit curves away from the observer. However, this proceeds more rapidly for the clouds on eccentric orbits, due to the torque experienced towards their off-plane pericentre passage as the leading end of the cloud arrives first and is pulled upwards (see Section~\ref{sec:maps}), which induces a slow clockwise rotation around the line of sight. As a result, the flattened clouds become inclined relative to the Galactic plane (see \autoref{fig:map}), which causes the aspect ratio along the $\{x,z\}$ coordinates to return to unity even when the clouds are still tidally flattened. This has a pronounced effect on the snapshot representing the Brick -- the cloud in \autoref{fig:map} is clearly flattened, with a minor-to-major axis ratio of $\sim0.3$, but is inclined relative to the Galactic plane by $\sim35\degr$, resulting in the elevated aspect ratio of $\delta z/\delta x\sim0.65$ in the third panel of \autoref{fig:struc} (indicated by the black square). Eventually, the flattening is erased altogether, most likely by turbulent mixing during the collapse of the clouds.

\subsubsection{Cloud column densities}

Finally, the bottom panel of \autoref{fig:struc} shows the evolution of the cloud column densities observed in the plane of the sky. We note that these densities represent simulations of single clouds and do not accurately reflect the stacked composite image of \autoref{fig:map}. With the exception of the high-density simulation, in which the column density decreases due to its rapid conversion into sink particles (see below), most clouds exhibit a similar evolutionary progression. Initially, the column density decreases due to the differential acceleration and shear along the orbit (which primarily takes place in the plane of the sky) and the corresponding longitudinal stretching of the clouds. However, eventually the compressive tidal field takes over, resulting in a significant (factor-of-several) increase of the column density that commences during the pericentre passage. By contrast, the models with circular orbits undergo a weaker column density increase ($<0.3~\dex$) that sets in $\sim0.1~\myr$ later. Near the position of the Brick, the column density reaches a local maximum, after which the cloud slowly expands in the latitudinal direction as it emerges from the bottom of the gravitational potential along its orbit. Due to this expansion, the global column density decreases slightly by 25--50~per~cent (naturally, this does not occur for the clouds evolving on circular orbits).

After the clouds reach the longitude of cloud e/f ($l\sim0\fdg5$), the column density monotonically increases. The cause for this increase is twofold. Firstly, the orbital curvature away from the observer geometrically compresses the projected dimensions of the clouds, because their radial extents at the end of the simulations are generally smaller than their azimuthal dimensions (cf.~\autoref{fig:map} and see \citealt{dale19}). The resulting orientation of the clouds along the line of sight is the main driver for the sharp upturn of the column density near the position of Sgr~B2. Secondly, the dynamical perturbation from the background potential weakens at larger galactocentric radii, allowing the velocity dispersion to decrease (see Section~\ref{sec:kin}) and gravitational collapse to set in.

In all simulated clouds, the most tightly bound clumps proceed to collapse and form sink particles at a high rate. Given the dimensions and column densities shown in \autoref{fig:struc}, the cloud-averaged free-fall time follows directly as $\tff\propto(\delta y/\Sigma)^{1/2}$ and for the clouds considered here ($\delta y\approx5~\pc$ and $\Sigma\approx2\times10^3~\msun~\pc^{-2}$) has typical values of $\tff\sim0.4~\myr$. Relative to the orbital dynamical time at pericentre $\Omega^{-1}=r_{\rm p}/v_{\rm orb,p}\approx0.3~\myr$ (which is the minimum along the orbit), this time-scale is comparable, implying that gravitational collapse proceeds rapidly once it has set in. While this rapid collapse is physical in nature, it leads to an unphysical bias towards a nearly monotonically decreasing column density in the high-density simulation. In the absence of feedback or magnetic fields, it is the only simulation to reach a star formation efficiency of 90~per~cent by the time the cloud reaches apocentre at the end of the simulation. The column densities of the other clouds are not significantly affected by the conversion of gas into sink particles, because the star formation efficiencies stay considerably lower, at 15--50~per~cent.

Combining the range of column densities spanned by all simulations, we see that any systematic evolutionary trends with Galactic longitude or time span a factor of 1.5--4 between the minimum column density and that at the end of the simulation. This is smaller than the total range of column densities spanned by the initial conditions, which is a factor of $\sim6$ and is representative for the variety of clouds observed upstream of the dust ridge (see Section~\ref{sec:model}). This difference in dynamic range means that, analogously to the cloud dimensions in the second panel of \autoref{fig:struc}, variations in column density along the dust ridge primarily trace variations in the initial conditions of the clouds. It also explains why observational studies have found only limited evidence for trends of the column density with Galactic longitude \citep[e.g.][]{krieger17}. A comparison between these results and the observed column densities is presented in Section~\ref{sec:comp}.

In summary, the morphological characteristics of the simulations all sketch a similar picture, across the full range of initial conditions. Driven by the background potential and the orbital eccentricity, the combination of geometric convergence during pericentre passage and the compressive tidal field result in a flattened and inclined, pancake-like cloud morphology and enhanced densities, eventually leading to fragmentation and gravitational collapse once the clouds have passed pericentre. The role of the pericentre passage is to accelerate this evolution relative to clouds evolving on circular orbits.

\subsection{Kinematics and dynamics} \label{sec:kin}

The kinematic evolution of the clouds is quantified in \autoref{fig:kin} for all five simulations. For reference, a colour composite of the inner CMZ with the orbital model highlighted is included as the top panel. The figure shows several key properties of the simulated clouds as a function of Galactic longitude, i.e.~their line-of-sight velocity dispersions, their line-of-sight velocity gradients in Galactic longitude, and their spin angular momenta. As before, the quantities are calculated for the gas with densities $\rho\geq10^4~\cmc$ (cf.~\autoref{fig:map}), except for the spin angular momentum (see below). The line-of-sight velocity dispersions ($\sigmalos$) are obtained by calculating the standard deviation of the line-of-sight velocities of the particles within the rectangle spanned by the one-dimensional half-mass extents, i.e.~$\{x_{\rm cm}\pm\delta x,z_{\rm cm}\pm\delta z\}$. This is equivalent to taking the mass-weighted average of the line-of-sight velocity dispersion at each coordinate $\sigmaloc(x,z)$ as
\be
\label{eq:sigma}
\sigmalos = \frac{\int_{x_{\rm cm}-\delta x}^{x_{\rm cm}+\delta x}\int_{z_{\rm cm}-\delta z}^{z_{\rm cm}+\delta z} \Sigma_{\rm local}(x,z)\sigmaloc(x,z) {\rm d}x{\rm d}z}{\int_{x_{\rm cm}-\delta x}^{x_{\rm cm}+\delta x}\int_{z_{\rm cm}-\delta z}^{z_{\rm cm}+\delta z} \Sigma_{\rm local}(x,z) {\rm d}x{\rm d}z} .
\ee
Likewise, we calculate the line-of-sight velocity gradient ($\dvdl$) by carrying out an orthogonal linear regression to the $\{l,\vlos\}$ distribution of the gas map within $\{x_{\rm cm}\pm\delta x,z_{\rm cm}\pm\delta z\}$ \citep{boggs90}. Again, this is equivalent to a mass-weighted fit to the position-velocity image. Finally, the spin angular momenta ($L_z$) are calculated without making any cuts in density or position, because this is a physical quantity used to describe the entire cloud and interpret the predicted velocity gradients, rather than an observable quantity. Given a population of $N_{\rm part}$ particles, we obtain
\be
\label{eq:angmom}
L_z = \sum_{N_{\rm part}}m_i(\mathbfit{r}_{xy,i}\times\mathbfit{v}_{xy,i}) ,
\ee
where $m_i$ is the particle mass, $\mathbfit{r}_{xy,i}$ is the position vector in the Galactic plane relative to the cloud's centre of mass, and $\mathbfit{v}_{xy,i}$ is the velocity vector in the Galactic plane relative to the cloud's centre of motion.

\subsubsection{Cloud line-of-sight velocity dispersions} \label{sec:sigmalos}

\begin{figure}
\includegraphics[width=\hsize]{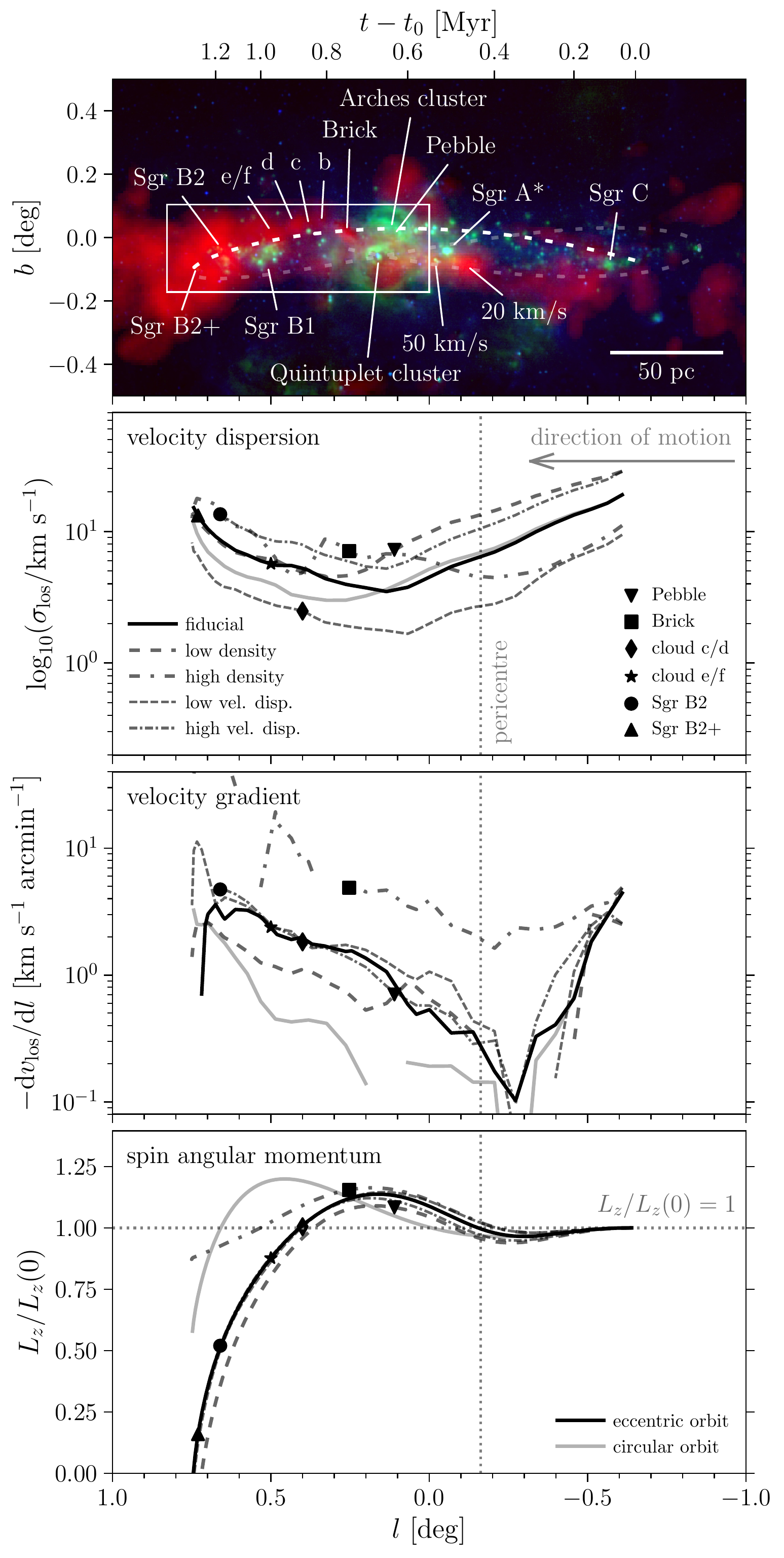}%
\vspace{-2.4mm}\caption{
\label{fig:kin}
Kinematic evolution of the five simulated clouds. Panel~1: three-colour composite image of the CMZ as in \autoref{fig:struc}. Panel~2: evolution of the cloud velocity dispersions along the line of sight. Panel~3: evolution of the cloud line-of-sight velocity gradients in Galactic longitude, i.e. $-\dvdl$. The lines are interrupted at longitudes where $-\dvdl$ is negative (see the text). Panel~4: evolution of the cloud spin angular momenta $L_z$, normalised to the initial $L_z(0)$. In all panels, the lines and symbols have the same meaning as in \autoref{fig:struc}, as indicated by the legends. This figure quantifies how the kinematic evolution of the clouds is shaped by their orbital motion through the gravitational potential of the CMZ.
}
\end{figure}

First focusing on the line-of-sight velocity dispersions (second panel of \autoref{fig:kin}), we see that the clouds initially lose internal kinetic energy, reach a minimum near the onset of the dust ridge at $l\approx0\fdg1$, and increase again towards Sgr~B2. The initial decline of $\sigmalos$ is not surprising, because we do not include any explicit turbulent energy driving in the simulations. Over time, the internal kinetic energy of the clouds should decrease. The only possible source of internal kinetic energy is the shear and associated tidal torques generated by the background potential, which effectively set a (cloud radius-dependent) velocity dispersion floor. Across the radii of the clouds shown in \autoref{fig:struc}, the shear-driven velocity differential across a cloud of scale $R_{\rm h}$, in a frame that orbits the Galactic centre with the cloud centroid, is given by\footnote{Because the velocity dispersion is a mass-weighted quantity, this expression requires the characteristic radius within the cloud where its mass resides, i.e.~the half-mass radius $R_{\rm h}$ rather than the truncation radius $R_{\rm t}$.}
\be
\label{eq:dv}
\begin{aligned}
\delta v&=(A+B) R_{\rm h}=-r\frac{{\rm d}\Omega}{{\rm d}r}R_{\rm h}=\frac{3-\alpha}{2}\Omega R_{\rm h} \\
&\approx 5.0~\kms~\left(\frac{\Omega}{1.7~\myr^{-1}}\right) \left(\frac{R_{\rm h}}{7.5~\pc}\right) ,
\end{aligned}
\ee
where $A$ and $B$ are the \citet{oort27} constants, $\Omega=1.7~\myr^{-1}$ is the average angular velocity of the orbital model \citep[Table~1 of][]{kruijssen15}, and $R_{\rm h}$ is normalised to a value typical for the simulated clouds at the onset of the dust ridge ($l\approx0\fdg1$), where $R_{\rm h}=3$--$12~\pc$. The resulting velocity differential is similar to the lowest velocity dispersions reached by the simulations, even to the extent that the minimum $\sigmalos$ of the clouds is linearly related to their radii (compare \autoref{fig:kin} to the average of $\delta x$ and $\delta z$ in \autoref{fig:struc}), as predicted by equation~(\ref{eq:dv}). This shows that cloud collapse and fragmentation from a shearing medium is significantly affected by galactic dynamics. The linear relation between the velocity differential $\delta v$ and cloud radius $R_{\rm h}$ may thus explain why the size-linewidth relation of CMZ clouds is steeper \citep[$\sigma\propto R^{0.7}$,][]{shetty12,kauffmann17} than the classical slope \citep[$\sigma\propto R^{0.5}$, e.g.][]{larson81,heyer15} found in the Galactic disc \citep[also see][]{dale19}. We intend to quantify this possibility further in a future paper.

The time-scale on which the minimum $\sigmalos$ is reached varies. If we assume that most of the velocity dispersion seen in the simulations reflects turbulent motion, the relevant time-scale is the turbulence dissipation time-scale, which in the absence of turbulence driving or other external perturbations is $\tdiss=R/\sigma$, with $R$ the scale on which the initial turbulence is driven \citep[e.g.][]{maclow04}. If instead the velocity dispersion reflects ordered motion, it is expected to evolve on a crossing time, which is also given by $t_{\rm cr}=R/\sigma$. In both expressions, we take $R$ to represent the initial cloud truncation radius $R_{\rm t}$ as in \autoref{tab:ics}, resulting in $\tdiss=t_{\rm cr}\approx0.56~\myr$ for all simulations. As expected, this is similar to the initial e-folding time of the velocity dispersions in \autoref{fig:kin}. Perhaps unsurprisingly in view of the varying external potential of our simulated clouds, the decay does not continue, but eventually a minimum $\sigmalos$ is reached. The time-scale required for reaching this minimum is shortest for the high-density simulation at $t-t_0\approx0.4~\myr$, which is similar to $\tdiss$ and also closely matches the cloud's initial free-fall time (see \autoref{tab:ics}). The other simulations reach their minimum $\sigmalos$ at around $t-t_0=0.6$--$0.8~\myr$. Given that the evolution of the cloud in the high-density simulation is largely decoupled from the background potential, as the cloud is undergoing collapse anyway (cf.~Section~\ref{sec:struc}), the correspondence between the minimum $\sigmalos$ and the initial free-fall time is to be expected. However, the evolution of the other clouds is affected more strongly by their interaction with the background potential, indicating that the additional internal kinetic energy driven by their motion in a shearing potential may be responsible for increasing $\sigmalos$.\looseness=-1

The above interpretation is underlined by the circular-orbit control experiments presented in Appendix~\ref{sec:circ} and \citet{dale19}. Clouds on circular orbits experience a similar evolution to the eccentric simulations discussed here, albeit somewhat more slowly -- as in Section~\ref{sec:struc}, we find that the pericentre passage accelerates the evolution of the clouds, driving an increase of their velocity dispersion in less than $0.2~\myr$ post-pericentre. However, our additional, isolated control runs that do not include an external potential at all reach their minimum $\sigmalos$ even more quickly \citep{dale19}, showing that the presence of shear is key in extending the clouds' resilience against internal kinetic (or turbulent) energy dissipation and collapse.

Once collapse eventually sets in, it leads to a clear increase of the line-of-sight velocity dispersion, which is primarily driven by the conversion of potential energy into internal kinetic energy \citep[as reported for simulations of isolated clouds by e.g.][]{vazquezsemadeni07,ibanezmejia16}. In our simulations, this happens at high Galactic longitudes, where the perceived increase of $\sigmalos$ is boosted further by orbital curvature. As quantified by \citet{dale19}, the azimuthal velocity dispersion (i.e.~along the orbital direction of motion) is enhanced by shear and typically exceeds the radial velocity dispersion. Towards high longitudes, the orbital motion becomes parallel to the line of sight and $\sigmalos$ becomes dominated by the azimuthal component of the velocity dispersion, thus adding to its increase. As in the discussion of the bottom panel in \autoref{fig:struc}, we note that these velocity dispersions represent simulations of single clouds and do not accurately reflect the stacked composite image of \autoref{fig:map}. The superposition of multiple clouds along the line of sight would lead to an even larger increase of the line-of-sight velocity dispersion than reported in \autoref{fig:kin}, which shows the velocity dispersions of the single clouds in each of our simulations.

Altogether, we see that our model makes the strong prediction that the line-of-sight velocity dispersion should increase steeply with longitude for $l\ga0\fdg5$. Given that the presented simulations are drawn from a representative set of initial conditions, the range in $\sigmalos$ spanned by the simulations provides a prediction for the velocity dispersions expected to be observed for real CMZ clouds. We reiterate that the superposition of multiple clouds along the line of sight may cause deviations upward of our predicted range of velocity dispersions. A comparison between these results and the observed velocity dispersions is presented in Section~\ref{sec:comp}.

\subsubsection{Cloud velocity gradients and spin angular momenta} \label{sec:spin}

The third panel of \autoref{fig:kin} shows the evolution of the line-of-sight velocity gradient in Galactic longitude ($-\dvdl$), which evolves substantially along the orbit. For all simulations, the velocity gradient generally has the opposite sign of the orbital rotation, in that $\vlos$ decreases with Galactic longitude. This is also evident in the bottom panel of \autoref{fig:map}, which shows that the line tracing the orbital motion in $\{l,\vlos\}$ space is nearly perpendicular to the $\{l,\vlos\}$ orientations of the individual clouds. We interpret this counter-gradient as the imprint of shear, which causes the inside of the cloud to overtake the outside, thus generating anti-clockwise rotation for the clockwise orbital motion modelled here.

Using equation~(\ref{eq:dv}), it is straightforward to estimate the magnitude of the shear-driven velocity gradient in the idealised case of a cloud in virial equilibrium without internal kinetic (or turbulent) energy dissipation. At the adopted distance of the Galactic Centre, $1\arcmin\approx2.41~\pc$, implying that we expect a typical shear-driven velocity gradient of $-\dvdl=1.6~\kms~\amin^{-1}$ or $-\dvdl=0.7~\kms~\pc^{-1}$. However, deviations from this gradient are expected in a dynamically evolving, turbulent cloud. Because the gradient is expected to be driven by shear, it manifests itself mainly as an azimuthal velocity differential between radially displaced mass elements. The resulting line-of-sight velocity gradient naturally vanishes close to $l=0\degr$, where the azimuthal motion is perpendicular to the line of sight. In addition, the pericentre passage on the eccentric orbits is accompanied by a vertical compression and torques (see Section~\ref{sec:maps}), which can cause the driven internal kinetic energy to be dissipated. Indeed, \autoref{fig:kin} shows that the velocity gradient at pericentre is weaker than the reference value estimated above. However, after the clouds climb out of the bottom of the gravitational potential and eventually undergo gravitational collapse, the conservation of angular momentum causes the clouds to increase their angular velocity, thus steepening the velocity gradient. This too can be seen in \autoref{fig:kin}, in which the velocity gradients steepen more strongly for the eccentric orbits than the circular-orbit control run and reach extreme values of $-\dvdl=3$--$10~\kms~\amin^{-1}$ near the position of Sgr~B2.

As shown by the full set of model lines in \autoref{fig:kin}, the initial velocity gradients are nearly indistinguishable, but quickly diverge during their subsequent evolution. The initial similarity of $\dvdl$ is by construction, because each of the simulated clouds has been initialised with a velocity field that is consistent with having formed out of the shearing medium, which effectively sets their initial angular velocity and thus their velocity gradient. For an incompressible cloud, the shear from the external potential would maintain this gradient during the simulation. However, the conservation of spin angular momentum (which is only satisfied to within 20~per~cent for $l\la0\fdg5$ due to tidal shear-driven torques, see below) during cloud collapse and the corresponding steepening of the velocity gradient implies a strong dependence on the initial cloud density. \autoref{fig:struc} shows that differences in the cloud density determine when collapse sets in, causing the high-density cloud to have become a factor of $\sim5$ smaller than the fiducial one by the end of the simulation. The same factor of $\sim5$ offset appears in the velocity gradient, as the high-density cloud has been spun up by its gravitational collapse.\footnote{The critical role of collapse in increasing the magnitude of the velocity gradient in the high-density simulation at $l\ga0\fdg2$ is underlined by the fact that the spin angular momentum decreases for these Galactic longitudes (again due to torques from the background potential, see below). This means that the evolution of the gradient must be dominated by a combination of change in the moment of inertia (i.e.~contraction) and viewing angle (i.e.~the azimuthal motion getting projected along the line of sight due to orbital curvature).} While all simulations exhibit some degree of counter-rotation, we thus find that the magnitude of the resulting velocity gradient is set by the initial volume density.

Due to the stochastic evolution of the clouds' substructure, the fitted velocity gradients occasionally flip sign, especially at the shallow gradients near $l=0\degr$ and in the high-density simulation, which is characterised by rapidly evolving substructure due to its short dynamical time. Over the course of the simulation, it is the only cloud in which the vast majority of the dense fragments end up accreting onto its centre of mass, causing its mass to grow while inducing major fluctuations of most quantities shown in \autoref{fig:struc} and \autoref{fig:kin}. In the velocity gradients, these brief episodes manifest themselves as interruptions of the lines.

The Brick is known to have a strong velocity gradient in the direction opposite to its orbital motion \citep{rathborne14}. Using the high-density simulation (which best represents the Brick, see \autoref{tab:snaps}), we carry out a linear regression to the position-velocity distribution to quantify this gradient in the simulation. \autoref{fig:map} shows that the cloud is lopsided, with the main gas concentration located off-centre at $\{l,y\}=\{0\fdg27,-65~\pc\}$. Centring the window on this concentration and increasing its width to $l=\{0\fdg303,0\fdg248\}$ such that it encloses the observed $\sim8~\pc$ size of the Brick (see Section~\ref{sec:brick}), we obtain a line-of-sight velocity gradient of $-\dvdl\approx5.9~\kms~\amin^{-1}$. This is similar to the observed velocity gradients reported in the literature, which range from $-\dvdl=7.43\pm0.34~\kms~\amin^{-1}$ \citep{rathborne14}\footnote{The value we provide here differs from that quoted by \citet{rathborne14}, because we correct a small conversion error. The gradient $-\dvdl=7.43\pm0.34~\kms~\amin^{-1}$ is obtained directly from the intensity-weighted velocity field of the Brick in the MALT90 HNCO~4(0,4)--3(0,3) emission provided by \citet{rathborne14}.} to $-\dvdl=9.50\pm0.34~\kms~\amin^{-1}$ \citep[where the error bars correspond to the propagated distance uncertainty]{federrath16}. Given that the high-density simulation reproduces the observed value to within $\sim25$~per~cent, we propose that the Brick's velocity gradient opposite to the orbital direction of motion is explained by shear. At least qualitatively, shear likely also explains why clouds e/f and the Sgr~B2 complex show signs of counter-rotation in their observed $\{l,\vlos\}$ distribution (see the bottom panel of \autoref{fig:map}).

Finally, the bottom panel of \autoref{fig:kin} illustrates the effect of shear during the clouds' pericentre passages on their spin angular momentum $L_z$. As discussed in Section~\ref{sec:model}, the initial velocity fields of the clouds were chosen to match the expected net rotation due to shear. However, the spin angular momenta in the simulations increase further, reaching a maximum at the position of the Brick ($l\sim0\fdg2$), briefly after pericentre. Up to the point of maximum $L_z$, the five simulations are nearly indistinguishable, indicating that the increase of the spin angular momentum is insensitive to the initial conditions. This is a natural result of the fact that the clouds all experience the same tidal shear-induced torques during their orbital motion and their pericentre passage, while the initial evolution of their moments of inertia is very similar (see the homologous radius evolution at these longitudes in \autoref{fig:struc}).

As the clouds recede from the Galactic Centre after pericentre passage, the spin angular momentum decreases again. However, it does so much more quickly than the increase towards the maximum and drops below the initial value of $L_z$. This is caused by the influence of tidal shear-driven torques on the clouds. The magnitudes of these torques (and hence the rate at which the spin angular momentum changes) are affected by cloud collapse, as illustrated by the difference between the five simulations. The fiducial and low/high velocity dispersion simulations are indistinguishable, because they have the same free-fall times (see \autoref{tab:ics}) and a largely homologous radius evolution. However, the low-density simulation (which collapses relatively slowly) loses its spin angular momentum even more quickly than the fiducial simulation, whereas the high-density simulation (which achieves collapse relatively quickly) loses spin angular momentum more slowly. This difference reflects the contrasting moments of inertia of the clouds. As the high-density simulation collapses more quickly, it is less affected by the torque driven by tidal shear. Finally, the cloud on a circular orbit does not experience an enhanced tidal torque at pericentre, retaining $L_z/L_z(0)>1$ some $0.3~\myr$ longer than the other clouds.

In summary, the kinematic and dynamical characteristics of the simulations all sketch a similar picture. Driven by the presence of the background potential, the associated tidal shear, and the resulting torques, the clouds exhibit net rotation and a sustained floor of their internal kinetic energies, part of which may be translated into turbulent motion. The corresponding spin angular momentum increases somewhat (by 20~per~cent) due to continued tidal shear-driven torques, which affect the clouds more rapidly due to the pericentre passage. Observationally, these processes manifest themselves through elevated velocity dispersions once gravitational collapse has set in (at $l\ga0\fdg5$), which at these longitudes are increased further by shear acting along the line of sight due to orbital curvature, and strong velocity gradients opposite to the orbital direction of motion. Indeed, both of these features are observed in the real CMZ, in the form of extreme velocity dispersions towards Sgr~B2 \citep[e.g.][]{henshaw16} and a strong velocity gradient in the Brick opposite to the orbital rotation \citep[e.g.][]{rathborne14,federrath16}.

\section{Discussion} \label{sec:disc}

\begin{table*}
 \centering
  \begin{minipage}{116mm}
\caption{Physical mechanisms in the model and their effect on the cloud dimensions ($\delta r$, $\delta\phi$, $\delta z$), aspect ratio ($\delta z/\delta\phi$), column density ($\Sigma$), velocity dispersion ($\sigmalos$), velocity gradient ($\dvdl$), and spin angular momenta around the radial ($L_r$) and vertical ($L_z$) directions.}
\label{tab:effects}
\begin{tabular} {@{}lccccccccc@{}}
  \hline
 Physical element & $\delta r$  & $\delta \phi$ & $\delta z$ & $\delta z/\delta\phi$ & $\Sigma$ & $\sigmalos$ & $\dvdl$ & $L_r$ & $L_z$ \\
  \hline
  Compressive tidal field & $-$ & $-$ & $-$ & $-$ & $+$ & $+$ & $+$ & $0$ & $0$ \\ 
  Shear (and associated torques) & $+$ & $+$ & $0$ & $-$ & $-$ & $+$ & $+$ & $0$ & $+/-$ \\ 
  Geometric convergence$^\star$ & $-$ & $+$ & $-$ & $-$ & $+$ & $+$ & $0$ & $0$ & $0$ \\ 
  Torque at pericentre$^\star$ & $0$ & $-$ & $+$ & $+$ & $0$ & $0$ & $0$ & $-$ & $0$ \\ 
  \hline
\end{tabular}\\
Note: The symbols ($-$, $0$, $+$) indicate the sign of the change induced by each physical mechanism on the variable in each column. Mechanisms unique to eccentric orbits are marked with a star. A positive change of the spin angular momentum implies anti-clockwise rotation when looking towards negative values of the specified axis. `Geometric convergence' includes the effect of `differential acceleration' towards pericentre. Because the listed physical mechanisms are all related to the background potential, we adopt a polar coordinate system centred on Sgr~A*. We encourage the reader interested in a particular element of this table to consult the discussion in Section~\ref{sec:quant}, which presents the projection of the polar coordinates relative to the line of sight due to orbital curvature, as well as the detailed time evolution of these quantities.
\end{minipage}
\end{table*}
This paper presents the structural and kinematic properties of five different numerical simulations of molecular clouds following the best-fitting eccentric orbit in the gravitational potential of the CMZ from \citet{kruijssen15}. It is found that the evolution of the clouds is closely coupled to the orbital dynamics. Specifically, their sizes, aspect ratios, column densities, velocity dispersions, line-of-sight velocity gradients, and spin angular momenta are demonstrated to be strongly influenced by the background potential and the pericentre passage. Following the discussion in Section~\ref{sec:quant}, we summarise in \autoref{tab:effects} how the various physical mechanisms included in our model affect the observables discussed in this work. In this section, we briefly discuss the strengths and weaknesses of the model, present a brief comparison to observations, and provide its key predictions for future observational work.

\subsection{Strengths and shortcomings of the model} \label{sec:fails}

The presented simulations capture the orbital and internal dynamics of the clouds, but omit several physical mechanisms that are potentially important, such as magnetic fields, cosmic rays, detailed chemistry, and stellar feedback. While this may obstruct drawing any conclusions concerning the absolute star formation rates and efficiencies of the clouds, it does allow us to isolate and characterise the morphological and dynamical evolution of the clouds, which is the main focus of this paper. Above all, this enables us to perform controlled experiments and obtain a systematic understanding of the interplay between cloud evolution and galactic dynamics in the CMZ, with implications for galactic centres in general.

Stellar feedback is not expected to strongly affect the structure and dynamics of the clouds prior to advanced gravitational collapse and widespread star formation, which is not achieved until the position of Sgr~B2. Therefore, it is a reasonable omission in the context of this work. Chemistry is an important ingredient when generating synthetic line emission maps for direct comparison to observations \citep[in the context of the CMZ, see e.g.][]{bertram16}. Clearly, a detailed chemical network and appropriate radiative transfer modelling enables the generation of considerably more realistic maps than the volume density limit of $\rho\geq10^4~\cmc$ that we adopted to create \autoref{fig:map}. However, we note that the CMZ clouds host notoriously complex chemistry, excitation conditions, and optical depth effects, to the extent that clouds have a different appearance in each molecular line \citep[e.g.][]{rathborne15}. Accurately modelling the molecular chemistry necessary for following the high-density gas tracers often used to observed CMZ clouds remains out of reach for the foreseeable future. Importantly, a considerable part of our analysis focuses on kinematics. Observational velocity measurements are less affected by molecular abundances or chemistry than integrated intensity measurements. Finally, magnetic fields have been suggested to affect the dynamics of CMZ clouds on sub-pc scales \citep[e.g.][]{pillai15}. However, this is based on the total magnetic field strength. In the absence of a coherent magnetic field, the unordered (turbulent) magnetic field mainly acts as a source of pressure \citep[for recent discussions, see e.g.][]{li11,pillai15,federrath16}. Estimates of the turbulent magnetic field strength yield Alfv\'{e}nic Mach numbers ${\cal M}_{\rm A}>1$ \citep{federrath16}, indicating that supersonic turbulence dominates the cloud structure. Our results substantiate this finding -- the fact that our simulations reproduce the observed spatial structure and kinematics of CMZ clouds, suggests that magnetic fields may not be dynamically important, but instead trace the turbulent flow in the CMZ clouds.

Another assumption worth discussing is the initial density profile of the clouds. The adopted Gaussian density profile does not represent a hydrostatic equilibrium solution, meaning that some part of the clouds' evolution may be caused by their initial progression towards equilibrium. Unfortunately, the presence of an external gravitational potential implies that even density profiles satisfying hydrostatic equilibrium in isolation, such as cored power laws \citep[e.g.][]{mckee07,keto10}, would still require an initial equilibration phase. To answer whether a significant part of the observed cloud evolution is caused by the choice of initial density profile, we therefore compare the simulations discussed here to isolated control runs in \citet{dale19},\footnote{Note that these control runs do not include any background potential and are distinct from the circular-orbit control runs discussed in Appendix~\ref{sec:circ}.} finding that the evolutionary trends presented in this work are unique to the clouds evolving in the background potential of the CMZ. Reassuringly, this shows that any influence of the initial density profile on the results is subdominant relative to the role of the external potential in governing cloud evolution.

A final caveat is the choice of orbital model. Other parameterisations or dynamical models for the 100-pc stream in the CMZ exist \citep[e.g.][]{sofue95,molinari11,ridley17}. However, out of all these models, the \citet{kruijssen15} model provides the closest match to the position-position-velocity structure of the dust ridge \citep[and the 100-pc stream in general, also see][]{henshaw16}, which motivates its use in this paper. Perhaps most importantly, many of the identified ways in which the orbital dynamics affect cloud evolution are not sensitive to the details of the orbital solution, but are set by the global properties of the gravitational potential. For instance, the sizes, aspect ratios, velocity dispersions, velocity gradients, and spin angular momenta are all most strongly affected by the instantaneous tidal field and shear, while carrying second-order imprints of the pericentre passage on the adopted orbital model. Therefore, we predict that many of our findings should also persist in alternative orbital geometries \citep[e.g.][]{sormani18}.

\subsection{Comparison to observed column densities and velocity dispersions} \label{sec:comp}

We now briefly compare the properties of the simulated clouds to those of the observed dust ridge clouds. The first of these comparisons is quantitative, as it considers the column densities and velocity dispersions as a function of position along the dust ridge as in \autoref{fig:struc} and \autoref{fig:kin}, mirroring the evolution with orbital time in our simulations. The second comparison follows in Section~\ref{sec:brick} and focuses on a single snapshot of our high-density simulations at the position of the Brick.

For the first of the above two comparisons, we derive observed column densities from the HiGAL cold dust map (Battersby et al.~in prep.; see the middle panel of \autoref{fig:map}) and velocity dispersions from the HNCO $4(0,4)-3(0,3)$ data obtained with the Mopra CMZ survey \citep{jones12}. \citet{henshaw16} selected this molecular line as their primary tracer of the gas kinematics, because HNCO is widespread in the CMZ, suffers minimally from self-absorption in high column density regions, and does not exhibit hyperfine structure.\footnote{Note that the latter two points minimise any spurious broadening of the spectral line profiles when fitting them with a Gaussian.} Due to its widespread nature, it likely traces the bulk of the gas in clouds \citep{jones12}, necessitating that we include all gas present in the simulations, instead of imposing a density cut as before. To ensure an appropriate comparison between the column densities and velocity dispersions of observed and simulated clouds, we subject them to a highly analogous analysis. The observations and simulations are first convolved to a common spatial resolution of $1\arcmin$. The column densities are then evaluated at a single $\{l,b\}$ coordinate per cloud. For the simulations, we select the coordinate with the highest column density within a $5\times5~\pc$ area centred on the cloud centre of mass. For the observations, this coordinate is taken to correspond to the cloud centre, which is identified by fitting a two-dimensional Gaussian to each cloud in the HiGAL column density map. The column densities are then obtained directly from the simulated and observed maps, at the pixel closest to the the selected coordinate. The uncertainty on the observed column densities is taken to be a factor of 2.

The velocity dispersions of the simulated clouds are measured by calculating the mass-weighted line-of-sight velocity dispersion within a square of $7.2~\pc$ (i.e.~$0\fdg05$) in width, again centred on the cloud centre of mass. This averaging scale matches that over which the observed velocity dispersions are extracted. When measuring the observed velocity dispersions, it is crucial to avoid contamination by the complex large-scale kinematic structure of the CMZ, with multiple streams intersecting along the line of sight \citep[e.g.][]{kruijssen15,henshaw16}. Therefore, we fit the HNCO spectra across the entire CMZ obtained using the spectral line fitting code {\sc SCOUSE} \citep{henshaw16}. For these fits, we adopt a spectral averaging area with a width of $7.2~\pc$ (i.e.~$0\fdg05$). We centre the spectral averaging areas on the cloud coordinates obtained above, reject velocity components unassociated with the clouds,\footnote{Specifically, we remove components with negative line-of-sight velocities relative to the local standard of rest, or velocities offset from the \citet{kruijssen15} orbit by more than $40~\kms$. We also exclude components containing less than 25~per~cent of the flux of the brightest component.} and calculate the intensity weighted average velocity dispersion of these components.
\begin{figure}
\includegraphics[width=\hsize]{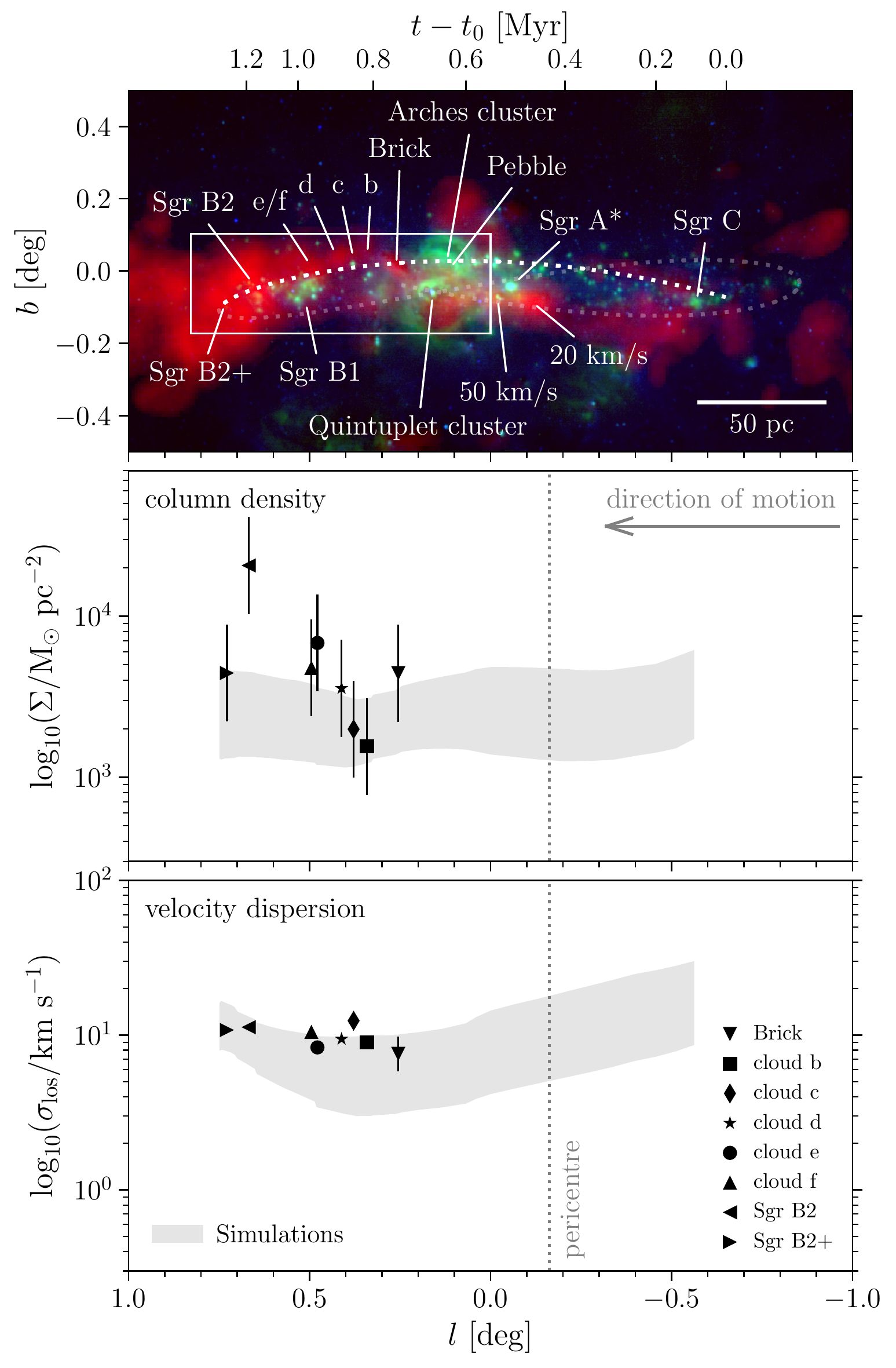}%
\vspace{-2mm}\caption{
\label{fig:obs}
Comparison of the simulations to observed dust ridge clouds. Panel~1: three-colour composite image of the CMZ as in \autoref{fig:struc}. Panel~2: column densities. Panel~3: line-of-sight velocity dispersions. The symbols represent the observed properties of the dust ridge clouds, whereas the grey-shaded area indicates the range spanned by the simulated clouds for the five different initial conditions from \autoref{tab:ics}. For the models, the column densities and velocity dispersions are calculated differently from those in \autoref{fig:struc} and \autoref{fig:kin} to facilitate an appropriate comparison to the observations (see the text). They are also averaged over a $0.3~\myr$ time window to reduce stochasticity. This figure shows that most of the observed column densities and velocity dispersions of dust ridge clouds are reproduced by drawing from the range of cloud properties observed upstream and simulating their evolution. Only the column density of Sgr~B2 exceeds the range covered by the simulations.
}
\end{figure}

The middle panel of \autoref{fig:obs} compares the observed column densities in the Brick, clouds~b, c, d, e, and f, Sgr~B2, and Sgr~B2+ to the total range spanned by the simulations (grey-shaded area) at each Galactic longitude or time. Neither the observations nor the simulations show any significant trends of increasing or decreasing column density with longitude (with the exception of the single data point marking Sgr~B2). For the simulations, we see that the total range spanned by the five different sets of initial conditions from \autoref{tab:ics} is wider than the amplitude of any trend across the full longitude range (as in \autoref{fig:struc}). The absolute column densities found in the simulations reproduce the observed range, with the exception of Sgr~B2. There are two possible reasons for this. Firstly, it may indicate the superposition of several clouds along the line of sight (as found in Section~\ref{sec:maps}). Secondly, our high-density simulation achieves an extreme star formation efficiency of nearly 90~per~cent near the position of Sgr~B2. This is unphysical and results from the absence of mechanisms that may slow down collapse and star formation, such as stellar feedback and magnetic fields. Including the mass locked in sink particles increases the upper envelope of the grey-shaded area by a factor of 5--10, making it consistent with Sgr~B2. Therefore, we conclude that the high column density of Sgr~B2 should be addressed in future simulations employing a more complete physical model.

In the bottom panel of \autoref{fig:obs}, we compare the observed velocity dispersions of the same set of clouds to the range covered in our simulations, again as a function of Galactic longitude. The observed velocity dispersions fall within the total range of simulated velocity dispersions. In addition, both the simulations and the observed clouds suggest a weak increase of $\sigmalos$ at $l>0\fdg2$, with Pearson and Spearman correlation coefficients for the observed data points of $r\approx0.5$. In the simulations, this is caused by a combination of (tidally-induced) gravitational collapse and shear (see Section~\ref{sec:sigmalos}). The generally good agreement at these longitudes contrasts with a slight ($2$--$3~\kms$) underprediction of the observed velocity dispersion of cloud~c. It is plausible that the difference results from differences in the initial conditions. If true, this would imply that cloud~c had initial conditions similar to the high-density and high-velocity dispersion simulations in \autoref{tab:ics}, which exhibit the highest velocity dispersions and define the upper bound on the grey-shaded area in the bottom panel of \autoref{fig:obs}. However, two important caveats are in order. Firstly, the observed velocity dispersions of CMZ clouds may vary by up to a factor of $\sim2$ depending on the spectral line used, even if they have comparable critical excitation densities \citep[e.g.][]{rathborne15}. With HNCO, we use a tracer showing widespread emission in the CMZ that likely traces most mass in clouds. It thus provides a good match to the simulations presented here, but it remains possible that other spectral lines are more appropriate. Secondly, we reiterate that the presented simulations capture a limited number of physical processes. The inclusion of additional physics, such as stellar feedback or chemistry, may add further scatter and widen the predicted ranges represented by the grey-shaded areas in \autoref{fig:obs}. We plan to address the influence of these processes in future work.

\begin{figure*}
\includegraphics[width=0.98\hsize]{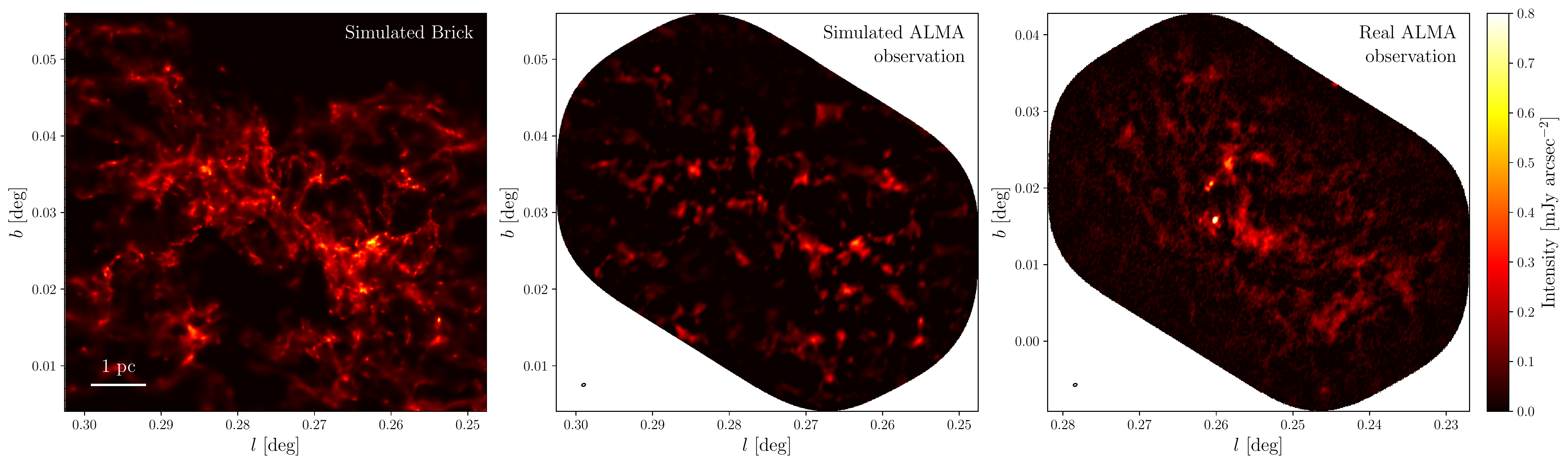}%
\vspace{-2mm}\caption{
\label{fig:brick}
Comparison of dust continuum observations of the `Brick' obtained with ALMA at 3~mm (89~GHz) to a simulated observation of the high-density simulation at the same position as the real Brick (cf.~\autoref{tab:snaps}). Panel~1: column density map of the native simulation, restricted to the high-density part of the gas reservoir associated with the cloud. Panel~2: simulated observation of panel~1, mimicking the precise setup of the ALMA observations (see the text). Panel~3: real ALMA observation from \citet{rathborne15}. All maps are scaled to the same units as indicated by the colour bar and described in the text. The ellipse in the bottom-left corners of panels~2 and~3 illustrates the beam shape and size. This figure shows that the morphology and typical peak brightness of the simulated cloud are similar to those of the observed Brick.
}
\end{figure*}

\subsection{Comparison to ALMA observations of the Brick} \label{sec:brick}

We now turn to a qualitative comparison of the observed 3~mm (89~GHz) dust continuum ALMA observations of the Brick \citep{rathborne15} to the high-density simulation snapshot at the position listed in \autoref{tab:snaps}, which provides the best match to the Brick among the simulations and snapshots presented in this work. To carry out this comparison, we generate a `simulated' ALMA observation using the \casa software package \citep{mcmullin07}, adopting the same setup with which the Brick was observed by \citet{rathborne15}. This is done using the {\tt simobserve} and {\tt simanalyze} tasks in {\sc casa}. These tasks allow us to perform synthetic ALMA observations of an input image by first simulating the observation based on user-defined input parameters and subsequently generating a set of corresponding visibilities. These visibility data are then imaged using the {\tt clean} task.

To prepare the simulated map, we first convert it from units of $\msun~\pc^{-2}$ to Jy. To do this, we assume a distance of $d=8.3~\kpc$ as in the rest of this paper, a dust temperature of $T=20~{\rm K}$, an observing frequency of $\nu=89~{\rm GHz}$, a gas-to-dust ratio of 100, and an emissivity index of $\beta=1.75$. The input parameters for this simulated observation are based on those presented in \citet[ALMA project ID 2011.0.00217.S]{rathborne15}. We use the {\tt buildConfigurationFile} task to generate antenna configuration files from the measurement sets of the real data. The data were taken across six execution blocks, each with varying antenna configurations. To replicate the observations, we simulate six separate observations, each corresponding to a different configuration. A mosaic of $72\arcsec\times162\arcsec$ with a central position of $\{l,b\} = \{0.275, 0.030\}$ is `observed',\footnote{Because the simulated cloud is lopsided (see \autoref{fig:map}), the central coordinates are offset from the centre of the Brick by $0\fdg026$. That way, the field of view is focused on where most of the mass resides in the simulation.} with four spectral bands set at 87.2, 89.1, 99.1, and 101.1~GHz, each with 1875~MHz bandwidth, thus providing a combined bandwidth of 7.5~GHz. Each execution is run for 40 minutes on-source and assumes a precipitable water vapour of ${\rm PWV}=1.5~{\rm mm}$.

\autoref{fig:brick} shows the original column density map of the high-density simulation at the position of the Brick, its simulated observation that has been generated with \casa according to the procedure above,\footnote{For consistency with most of the analysis presented in this work, we have used the same density threshold of $\rho\geq10^4~\cmc$ to produce the images of the simulation. Material at lower densities is unlikely to contribute strongly to the ALMA maps, because emission on large angular scales is filtered out by the interferometer.} and the dust continuum ALMA image of the real Brick from \citet{rathborne15}, all on the same (linear) intensity scale indicated by the colour bar. This comparison provides several relevant insights. Firstly, the setup used to observe the Brick with ALMA only recovers part of the flux and structure from the simulation. Relative to the left-hand panel of \autoref{fig:brick}, the middle panel especially misses emission near the edges of the field of view, where the sensitivity drops. However, the brightest cores and filamentary structures are recovered well. A qualitative\footnote{We reiterate that a one-to-one, absolute comparison between the simulated and observed maps is not meaningful, because the precise structure of the simulated map depends entirely on the specific realisation of the simulation's initial conditions. Even between identical realisations of the initial conditions, differences will develop during the simulation due to micro-scale chaos and stochasticity \citep[e.g.][]{keller19}.} comparison of these to the true Brick dust column density map in the right-hand panel shows remarkable agreement. Other than the observed core at $\{l,b\}=\{0\fdg261,0\fdg016\}$, which is known to host a water maser marking the onset of star formation and is saturated on the colour scale of \autoref{fig:brick} at $2.4~{\rm mJy}~\asec^{-2}$, the brightest cores in both the simulated and real Brick are of the order $0.5~{\rm mJy}~\asec^{-2}$. In addition, both maps show core sizes of $0.1$--$0.2~\pc$, connected by pc-scale, flocculent filamentary structures. Globally, they also follow a similar morphology, which manifests itself in the form of comparable inclinations, aspect ratios, and curvature. In both maps, the filamentary structures run along the major axis near the middle of the cloud (which we attribute to the vertical tidal compression in Section~\ref{sec:maps}) and fan out to cover the full width of the cloud towards its extremities in the top left and bottom right.

The above, cursory comparison of the simulated and real Brick clouds will be followed up with a more detailed analysis in future work. For instance, the size-linewidth relation, spatial and velocity power spectra, fractal dimension, and column density PDF are ideally suited observables for a thorough, quantitative comparison of these simulations to the observed CMZ clouds. In particular, an ALMA Large Programme covering the entire CMZ with a spatial resolution, sensitivity, and spectral setup as in \autoref{fig:brick} is both observationally feasible and would enable systematic comparisons of these quantities as a function of (orbital) position to the simulations.

\subsection{Predictions for future observational tests} \label{sec:pred}

Throughout this paper, we have drawn a direct comparison between the simulated clouds and observations of real CMZ clouds. This comparison shows that the presented simulations quantitatively reproduce a surprising variety of properties of the observed dust ridge clouds, from their column densities and velocity dispersions (Section~\ref{sec:comp}) to the velocity gradient (Section~\ref{sec:spin}) and spatially-resolved structure (Section~\ref{sec:brick}) of the Brick. These results add to the qualitative discussion of a much wider variety of observables in Section~\ref{sec:maps} and justify the use of these simulations as interpretative tools for constraining which physical mechanisms govern the baryon cycle in the CMZ. The quantitative character of these simulations opens up a variety of future observational tests of the physical predictions and hypotheses put forward in this work. Here, we focus on predictions of our models that would benefit from additional observational follow-up work. In addition, we discuss how broadly the predictions of our models should apply.

In principle, the physical processes modelled in this work should apply generally. We have adopted initial conditions representative of the clouds some $0.75~\myr$ upstream of the dust ridge, situated on the other side of the preceding pericentre passage, at negative Galactic longitudes and positive Galactic latitudes \citep[cf.][]{henshaw16b}. Based on their relatively low densities relative to other CMZ clouds, these clouds are assumed to have recently condensed out of the diffuse medium, possibly due to their recent arrival on the 100-pc stream. If the progenitors to other dense, dust ridge-like gas structures have similar properties, we would expect that these sequences of dense gas clouds also follow the behaviour found in our simulations. In this context, specific targets of interest could be the $20$ and $50~\kms$ clouds. These clouds are passing through pericentre at a radius of $r=60$--$70~\pc$ \citep{kruijssen15}, represent the high-density end of a part of the 100-pc stream that starts more diffusely near Sgr~C, and exhibit signs of star formation \citep[e.g.][]{ho85,mills11,lu15,lu17}. It is quite plausible that the morphological and kinematic trends with longitude identified in this work are found not only in the dust ridge, but at least qualitatively also in the $20$ and $50~\kms$ clouds.

We emphasise that the trends with Galactic longitude (or equivalently with time relative to accretion or pericentre passage) predicted by \autoref{fig:struc} and \autoref{fig:kin} are not necessarily monotonic. The range of initial conditions was chosen to represent a reasonable range based on the plausible precursor clouds to the dust ridge. As such, the spread between the model curves in \autoref{fig:struc} and \autoref{fig:kin} provides an uncertainty range. For some quantities (e.g.~the column density and velocity dispersion, see Section~\ref{sec:comp}), this range exceeds the total magnitude of the trend, implying that even opposite trends may be observed if the initial conditions of the clouds vary systematically with longitude. Notwithstanding this caveat, we do predict a weak, but measurable trend of increasing velocity dispersions towards the highest longitudes, which is tentatively confirmed by the observations shown in \autoref{fig:obs} (also see \citealt{krieger17}). Other quantities (e.g.~the aspect ratio and spin angular momentum) exhibit well-defined trends that represent firm predictions of this work. A systematic study of the presented observables as a function of position along the 100-pc stream should allow recently-condensed sequences of clouds to be identified. These sequences may cover any range of evolutionary phases between the initial condensation out of the diffuse medium and their eventual disruption by stellar feedback. However, the finite lengths of such correlated segments mean they are unlikely to span the complete evolutionary timeline at a single moment in time. Correlating the occurrence of these sequences with pericentre passages will show whether the eccentricity of the orbit plays a defining role in regulating cloud morphology in kinematics or, as expected based on these simulations, plays a relevant yet sub-dominant role next to the presence of the external gravitational potential.

Finally, the advent of ALMA now enables extragalactic CMZs to be observed at a sensitivity and spatial resolution similar to those obtained of the Galactic CMZ in the pre-ALMA era. Provided that the gravitational potential, as parameterised through the power law slope of the enclosed mass profile, and the orbital eccentricity of the gas streams in such CMZs are similar to those adopted here, our predictions should apply directly to these systems too. Discussed observables that may feasibly be obtained in face-on CMZs are the clouds' longitudinal extents ($\delta x$) and velocity dispersions ($\sigmalos$, under the assumption that the trends with longitude are stronger than the deviations from velocity isotropy). In addition, galaxies with low inclinations provide a unique opportunity to quantify the extents of clouds in the galactic plane, which we predict to be substantial. The predictions for the cloud aspect ratios, column densities, and velocity gradients are specific to observations through the orbital plane. Highly-inclined, edge-on galaxies such as NGC~253 \citep[e.g.][]{sakamoto11} would enable a comparison to these observables, in addition to the clouds' longitudinal extents and velocity dispersions.

Above all, the simulations presented in this paper constitute a rich data set that can be used to shed light on a wide range of open questions. While we have considered several key quantities and observables here, future work will extend our analysis to e.g.~the star formation efficiencies, virial parameters, and size-linewidth relations. Together with its companion paper \citep{dale19}, the present work sets the first step towards an in-depth physical understanding of the close interaction between galactic dynamics, cloud evolution, and star formation, both in the CMZ of the Milky Way and in extragalactic centres.

\section{Conclusions and implications} \label{sec:concl}

We have presented a set of five numerical simulations of gas clouds orbiting on the 100-pc stream of the CMZ, spanning a representative range of initial conditions, with the goal of characterising their morphological and kinematic evolution in response to the external gravitational potential and their passage through pericentre.\footnote{A detailed discussion of the simulations is presented in a companion paper \citep{dale19}, in which we also discuss the general properties of clouds orbiting in external potentials and investigate how the background potential affects the star formation activity of the orbiting clouds.} This represents the first set of numerical simulations specifically aimed at modelling the clouds in the CMZ dust ridge, thus enabling direct comparisons to observations. Indeed, we find that the inclusion of the background potential and the orbital motion allow our models to reproduce several key features of the observed CMZ clouds. The main results of this work are as follows.
\begin{enumerate}
\item[{\it Background potential}:]
The presence of a background potential and the clouds' motion through pericentre represent a transformational event, affecting several of the cloud properties. The potential generates a fully compressive tidal field in the galactocentric radius range $45<r/\pc<115$, most strongly so in the vertical direction, and also imposes a significant amount of shear. The pericentre passage increases the strengths of both these effects by several tens of per~cent. Additionally, the motion from apocentre towards pericentre on an eccentric orbit causes a geometric compression perpendicular to the direction of motion due to the convergence of the orbital trajectories, as well as a geometric extension along the orbit due to differential acceleration.
\item[{\it Global morphology}:]
The combination of the compressive tidal field and geometric convergence towards pericentre causes the clouds to be compressed vertically, leading to a factor 2 decrease of their vertical extent. In addition, the presence of shear and differential acceleration along the orbit stretches the clouds in the Galactic longitude direction by up to a factor of 2. Together, these effects turn the clouds into pancake-like structures, reaching an extreme aspect ratio of $\delta z/\delta x=0.25$ between pericentre and the position of the Brick, after which the aspect ratio gradually returns to unity by the position of Sgr~B2. These galactic-dynamical deformations affect the simulated clouds down to the scales below which self-gravity dominates, causing them to dynamically decouple from the background potential. Quantitatively, the decoupling scale varies with the cloud density.
\item[{\it Column densities}:]
Due to the vertical compression, the (mass-weighted) column densities of the simulated clouds reach a local maximum at the position of the Brick. The global column density subsequently briefly decreases, but the densities of individual sub-clumps proceed to increase as the clouds emerge from the bottom of the potential at pericentre and gravitational collapse sets in. The curvature of the orbit away from the observer projects the azimuthally-stretched clouds along the line of sight, driving up the column density. However, the combined magnitude of these evolutionary trends is smaller than the range of column densities encompassed by our suite of simulations, which were chosen to be representative of the gas upstream from the dust ridge. Therefore, column density variations among the dust ridge clouds primarily trace variations in their initial conditions. We also find that the observed rise of the global column density towards high Galactic longitudes at the Sgr~B2 complex indicates the superposition of multiple clouds along the line of sight.
\item[{\it Spatial structure}:]
As the clouds orbit the Galactic Centre, their central regions fragment and undergo local gravitational collapse, whereas their outer layers disperse under the influence of shear. This causes the clouds to develop flocculent filamentary structures oriented perpendicularly to the vertical compression, and form $10~\pc$-scale extensions pointing towards the observer. Due to the torque experienced by the clouds as they pass through pericentre, these extensions are offset towards high Galactic longitudes and latitudes, resulting in clouds that appear inclined in the plane of the sky. Some of these extensions are (partially) accreted by the cloud centres over time, causing them to grow in mass. This more strongly affects the higher-density clouds.
\item[{\it Kinematics and dynamics}:]
The kinematics of the clouds are driven by a combination of shear and gravitational collapse due to the compressive tidal field. The dissipation of the initial turbulent energy and the corresponding decrease of the line-of-sight velocity dispersion are slowed down by shear-generated motions. These motions cause the clouds to counter-rotate relative to the orbital motion, turning them into spinning pancakes with velocity gradients opposite to the orbital rotation. Eventually, gravitational collapse sets in and causes the velocity dispersions to increase. Our models make a strong prediction that the velocity dispersion should increase steeply with longitude between $0\fdg5\leq l\leq0\fdg8$. As the orbit curves off at high longitudes, the superposition of clouds along the line of sight leads to extreme kinematic complexity, with large numbers of velocity components. Another consequence of the collapse is that the clouds' velocity gradients increase in strength between pericentre passage and the position of Sgr~B2 due to the collapse-driven increase of their angular velocity. The clouds' spin angular momenta are not conserved during this collapse. They reach a peak value at the position of the Brick due to the preceding pericentre passage, but decrease during collapse due to tidal shear-driven torques from the background potential.
\item[{\it Comparison to observations}:]
The above quantities and their evolution naturally explain a number of key observations of CMZ clouds. Below, we list the main ones, indicate the responsible mechanisms in parentheses, and provide key observational references where appropriate. The simulations reproduce the Brick's high column density (compressive tidal field and pericentre passage, cf.~\citealt{longmore12}), its velocity gradient opposite to the orbital rotation (shear, enhanced by collapse due to the compressive tidal field, cf.~\citealt{rathborne14,federrath16}), its flattened morphology (compressive tidal field and geometric deformation during pericentre passage, cf.~\citealt{lis98}), its inclination in the plane of the sky (torque during pericentre passage), its expanding outer layers (shear, cf.~\citealt{rathborne14}), and its filamentary structure along its major axis (compressive tidal field, cf.~\citealt{rathborne15}; also seen in clouds b, d, and e by \citealt{walker18}). They also reproduce the evolution along the dust ridge of the clouds' inclination angles (torque during pericentre passage), as well as the increase of the velocity dispersion along the dust ridge towards Sgr~B2 (compressive tidal field and shear, cf.~\citealt{henshaw16}), the increased kinematic complexity of the Sgr~B2 complex (shear and orbital curvature, cf.~\citealt{mehringer93}), and the range of cloud column densities found along the dust ridge (initial conditions, compressive tidal field, shear, orbital curvature, and pericentre passage, cf.~Battersby et al.~in prep.). Taken together, the reproduction of such a wide range of observables strongly suggests that the dynamical ingredients of the presented models are critical for understanding the properties, formation, and evolution of the CMZ clouds.
\item[{\it Background potential vs.~pericentre passage}:]
\citet{dale19} compare the simulations discussed here to a set of control experiments of clouds on a circular orbit within the same gravitational potential (thus `switching off' the pericentre passage), as well as of clouds in complete isolation. We provide analogues to \autoref{fig:struc} and \autoref{fig:kin} for the circular control runs in Appendix~\ref{sec:circ}, but refer to \citet{dale19} for further details and summarise the conclusions relevant to the results of this work here. We find that the presence of the background potential is the main factor in setting the behaviour of most of the observables discussed above. The potential generates shear and shapes the tidal field, thereby setting the cloud radii, aspect ratios, the onset of gravitational collapse and the corresponding rise of the column densities, velocity dispersions, velocity gradients, and absolute spin angular momenta. However, the pericentre passage accelerates the evolution of all of the above observables by $0.1$--$0.3~\myr$ and additionally drives the temporary decrease of the column densities downstream from the Brick due to post-pericentre expansion, as well as the rise of the spin angular momenta just after pericentre. The extent to which these effects manifest themselves in practice depends on the timing of pericentre passage relative to cloud condensation or accretion (see below).
\item[{\it Collapse triggered by pericentre passage:}]
As a result of the above comparison between the background potential and the pericentre passage, we find that the pericentre passage may act as a trigger for collapse (and possibly star formation) if the gas enters the 100-pc stream (either by accretion or by condensing out of the diffuse medium) less than one free-fall time before pericentre (i.e.~$\Delta t\la0.5~\myr$ or $\Delta l\la0\fdg5$ for the gas upstream from the dust ridge, see \citealt{henshaw16b}). This can manifest itself as an evolutionary progression of clouds as a function of Galactic longitude downstream from pericentre. However, if the gas enters the stream earlier, it may collapse without the aid of the `nudge' provided by the pericentre passage, thus interrupting any evolutionary sequence. Note that the apocentre radius of the \citet{kruijssen15} orbit ($r_{\rm a}=120~\pc$) lies on the outer edge of the compressive region ($45<r/\pc<115$), which implies that clouds orbiting on the gas stream may alternate between mildly extensive and strongly compressive tidal fields. Given the time difference between apocentre and pericentre of $\Delta t=1~\myr$, any self-gravitating gas with volume densities $\rho\la10^3~\cmc$ (corresponding to free-fall times $\tff\ga1.1~\myr$) will thus undergo collapse triggered by the pericentre passage. For higher-density gas, this depends on the time of cloud condensation or its accretion onto the 100-pc stream relative to pericentre.
\item[{\it Collapse triggered by accretion onto the 100-pc stream:}]
We find that the strong influence of the compressive tidal field causes major changes to the evolution of clouds orbiting in the CMZ potential. Any gas flows that enter the compressive region at $45<r/\pc<115$ will be subject to the sudden presence of a fully compressive tidal field and are therefore likely to follow the evolutionary streamline identified in this paper. The subsequent evolution of the condensing clouds should follow the general predictions made by our simulations. This means that evolutionary sequences of CMZ clouds may follow from a preceding pericentre passage if the timing is right, but are not restricted to such `hotspots'. Instead, any evolutionary progression found among segments of the 100-pc stream may be used to infer sites of cloud condensation or accretion upstream from such segments and can plausibly be translated into an absolute timeline by using the simulations presented here. This channel for collapse is most important for clouds that condense or accrete onto the 100-pc stream more than a free-fall time before the next pericentre passage. We thus expect evolutionary sequences along the 100-pc stream to appear segmented, with a fraction of them being triggered by pericentre passage and another fraction being triggered by entering the stream.\looseness=-1
\item[{\it General implications}:]
The results presented in this paper reveal that the evolution of molecular clouds near galactic centres is closely coupled to their orbital dynamics. As galactic rotation curves must turn over at small radii, such that $v\propto r^{(\alpha-1)/2}\equiv r^\beta$ with $\alpha>2$ and $\beta>1/2$, a fully compressive tidal field is predicted to be present in most galactic centres. As a result, the accretion of gas onto these galactic centres will be accompanied by transformative dynamical changes to the clouds, which likely lead to their collapse and associated star formation. During their subsequent evolution, the clouds are shaped by high levels of shear, as well as tidal and geometric deformation. Together, these processes naturally give rise to the starbursts observed in numerous galactic nuclei \citep[e.g.][also see \citealt{krumholz15,krumholz17,torrey17}]{jogee05,davies07,leroy13}.
\end{enumerate}
By zooming in on individual, simulated clouds that orbit in a realistic gravitational potential of a galactic nucleus, we have identified several key physical processes that govern the lifecycle of gas and star formation in such nuclei. It is not unlikely that these mechanisms set important bottlenecks (or avenues) for gas accretion onto supermassive black holes, thus affecting the large-scale evolution of the host galaxy. For instance, the tidally-induced, efficient circumnuclear star formation may represent an important accretion bottleneck and explain why there is no correlation between supermassive black hole growth and the presence of galactic bars \citep[e.g.][]{goulding17}. In the future, it will be beneficial to expand our models to a broader range of spatial scales and observables, to further increase their predictive power and facilitate additional, direct comparisons to the Galactic CMZ and extragalactic nuclei.

\section*{Acknowledgements}

JMDK and SMRJ gratefully acknowledge funding from the German Research Foundation (DFG) in the form of an Emmy Noether Research Group (grant number KR4801/1-1). JMDK and MAP gratefully acknowledge funding from the European Research Council (ERC) under the European Union's Horizon 2020 research and innovation programme via the ERC Starting Grant MUSTANG (grant agreement number 714907) and from Sonderforschungsbereich SFB 881 ``The Milky Way System'' (subproject P1) of the DFG.
ATB gratefully acknowledges funding from the ERC under the European Union's Horizon 2020 research and innovation programme (grant agreement number 726384). 
ASM and STS acknowledge funding by the DFG via the Sonderforschungsbereich SFB 956 ``Conditions and Impact of Star Formation'' (subproject A4 and A6) and the Bonn-Cologne Graduate School.
This work has made use of the Python libraries Matplotlib \citep{hunter07}, Numpy \citep{vanderwalt11}, Scipy \citep{jones01}, and Astropy \citep{astropy13}.
We thank Christoph Federrath, Simon Glover, Ralf Klessen, Jill Rathborne, and Mattia Sormani for insightful comments and discussions during the development of this work, and Henrik Beuther, Jens Kauffmann, and Peter Schilke for their helpful feedback on an early draft. We thank an anonymous referee for constructive comments that improved this paper. JMDK and SNL thank Schlo\ss\ Weitenburg for their kind hospitality during the development of this work.

\bibliographystyle{mnras}
\bibliography{mybib}

\section*{Supporting information}

Supplementary data are available at MNRAS online.\\[1ex]
\noindent \textbf{Movie.} Evolution of the simulated dust ridge clouds: animated version of \autoref{fig:map} showing the full evolutionary time sequence.\\[1ex]
\noindent Please note: Oxford University Press is not responsible for the content or functionality of any supporting materials supplied by the authors. Any queries (other than missing material) should be directed to the corresponding author for the article.

\appendix

\section{Structural and kinematic evolution in the circular control runs} \label{sec:circ}
\begin{figure}
\includegraphics[width=\hsize]{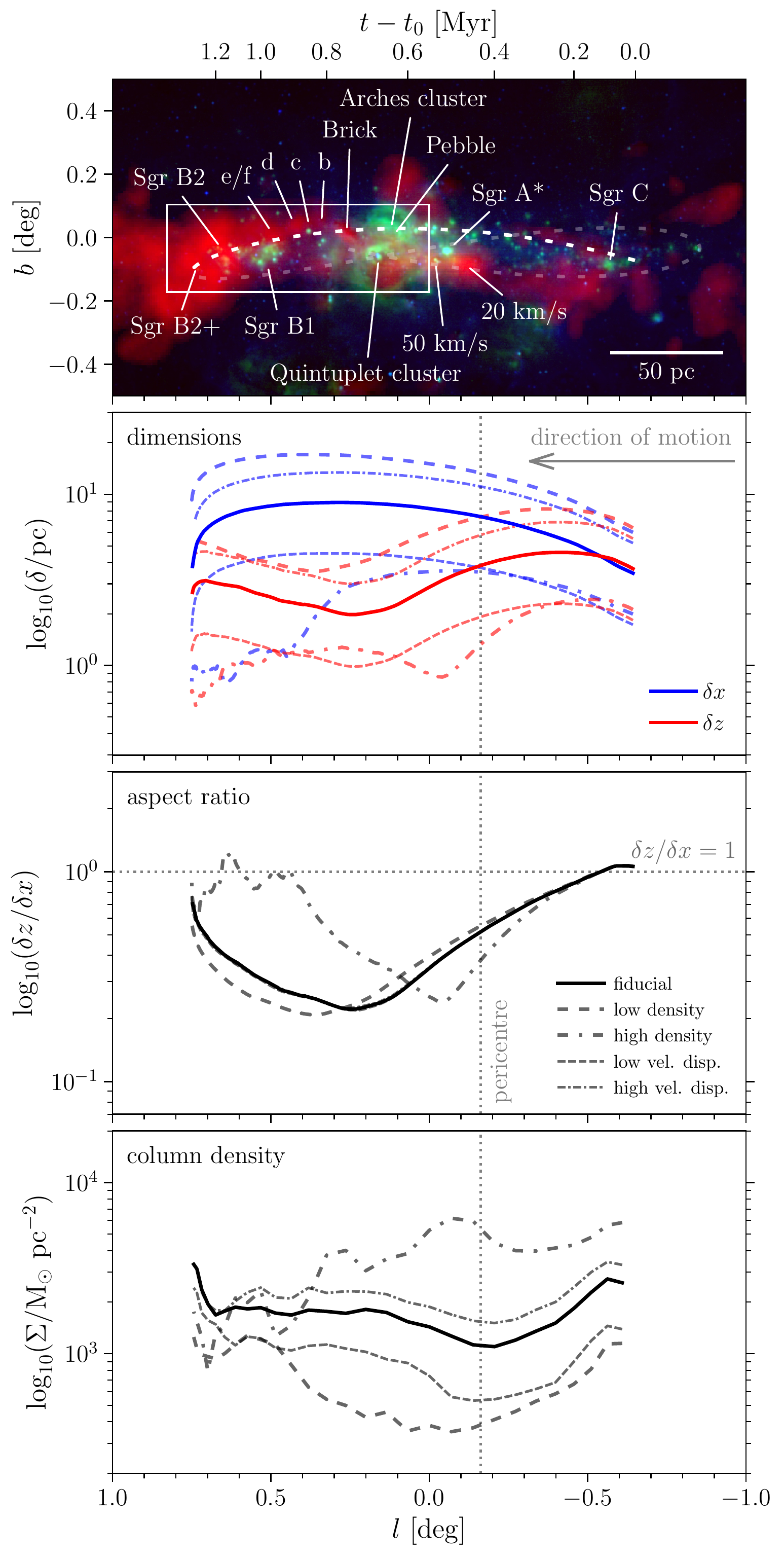}%
\vspace{-2mm}\caption{
\label{fig:struc_circ}
Morphological evolution of the five simulated clouds on circular rather than eccentric orbits. Panel~1: three-colour composite image of the CMZ as in \autoref{fig:struc}. Panel~2: evolution of the cloud dimensions, represented by their longitudinal ($\delta x$, blue) and latitudinal ($\delta z$, red) half-mass radii. Panel~3: evolution of the cloud aspect ratios, i.e. $\delta z/\delta x$. Panel~4: evolution of the mass-weighted cloud column densities. In all panels, the lines refer to different initial conditions in the same way as \autoref{fig:struc} (see the legend).
}
\end{figure}

\begin{figure}
\includegraphics[width=\hsize]{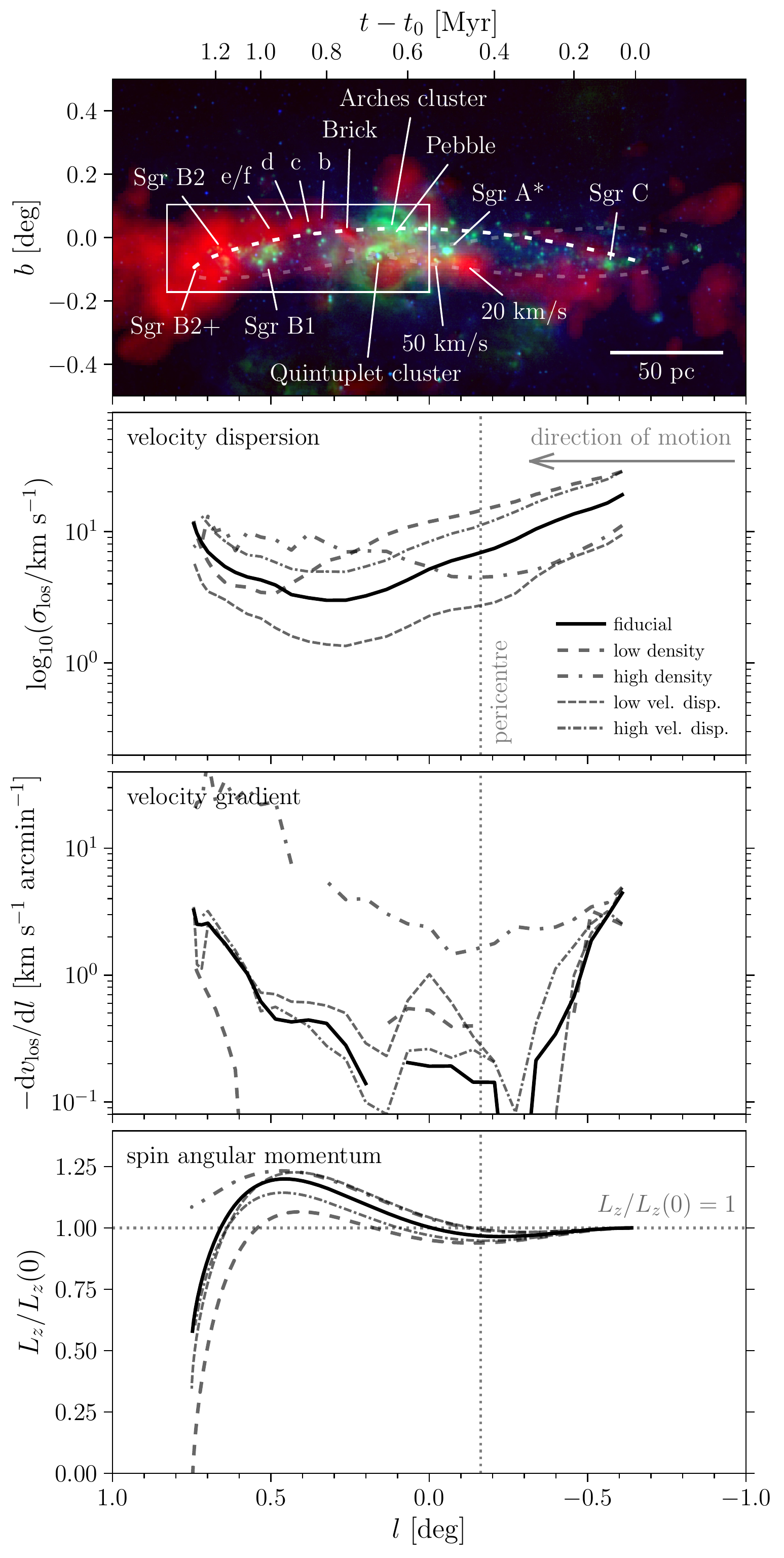}%
\vspace{-2mm}\caption{
\label{fig:kin_circ}
Kinematic evolution of the five simulated clouds on circular rather than eccentric orbits. Panel~1: three-colour composite image of the CMZ as in \autoref{fig:struc}. Panel~2: evolution of the cloud velocity dispersions along the line of sight. Panel~3: evolution of the cloud line-of-sight velocity gradients in Galactic longitude, i.e. $-\dvdl$. The lines are interrupted where $-\dvdl$ is negative. Panel~4: evolution of the cloud spin angular momenta $L_z$, normalised to the initial $L_z(0)$. In all panels, the lines refer to different initial conditions in the same way as \autoref{fig:kin} (see the legend).
}
\end{figure}

For reference, \autoref{fig:struc_circ} and \autoref{fig:kin_circ} quantify the evolution of the simulated clouds on circular orbits, analogously to \autoref{fig:struc} and \autoref{fig:kin}, respectively. We have chosen the radius of these circular orbits to match the initial galactocentric radius of the clouds on eccentric orbits, i.e.~$r=90~\pc$ \citep{dale19}. This radius represents the mean galactocentric radius of the eccentric orbits (which range from $r=60~\pc$ to $r=120~\pc$) and also provides a close match to their time-averaged radius ($r=97.3~\pc$). It thus represents the tidal environment in which the clouds on eccentric orbits spend most of their time. Overall, the main difference between clouds evolving on these orbits is that the pericentre passage of the eccentric orbit accelerates cloud evolution by $0.1$--$0.3~\myr$ relative to clouds on circular orbits. See Section~\ref{sec:quant} for further details.

\clearpage

\bsp

\label{lastpage}

\end{document}